\documentclass{ws-jai}
\usepackage{booktabs}
\usepackage{xcolor}
\usepackage{microtype}
\usepackage{placeins}
\usepackage{float}

\usepackage{needspace}
\pdfpagewidth=\paperwidth
\pdfpageheight=\paperheight
\renewcommand{\draftnote}{}
\usepackage[hidelinks]{hyperref}
\makeatletter
\def\toclevel@appendix{1}
\def\toclevel@subappendix{2}
\makeatother
\renewcommand{\theappendixc}{\Alph{appendixc}}
\let\JAIappendix\appendix
\renewcommand{\appendix}[1]{\Needspace{8\baselineskip}\JAIappendix{#1}}
\let\JAIsubappendix\subappendix
\renewcommand{\subappendix}[1]{\Needspace{7\baselineskip}\JAIsubappendix{#1}}
\hypersetup{pdftitle={SimpleClock: a satellite-disciplined oscillator for astronomical instrumentation},pdfauthor={Albert Wai Kit Lau}}

\begin{document}
\raggedbottom
\catchline{}{}{}{}{}
\makeatletter
\def\@clinebuf{\catchlinefont\noindent Manuscript prepared for Journal of Astronomical Instrumentation\par\noindent Draft manuscript\par}
\makeatother
\markboth{A. W. K. Lau et al.}{SimpleClock: a satellite-disciplined oscillator}
\title{SimpleClock: a satellite-disciplined oscillator\\for astronomical instrumentation}
\author{Albert Wai Kit Lau$^{1}$, Bruno Pollarolo$^{2}$, Juan Mena-Parra$^{1,2}$}
\address{$^{1}$Dunlap Institute for Astronomy and Astrophysics, University of Toronto, Toronto, Canada\\
$^{2}$David A. Dunlap Department of Astronomy and Astrophysics, University of Toronto, Toronto, Canada}
\maketitle
\begin{abstract}
SimpleClock is a low-cost, high-stability timing reference designed for astronomical experiments that combines a global navigation satellite system (GNSS) timing receiver, a digitally tunable 10 MHz microelectromechanical systems (MEMS) oven-controlled oscillator, external thermal regulation and autonomous control. It provides 10 MHz and pulse-per-second references at an estimated prototype hardware cost of US\$500. Independently enabled frequency and phase feedback support precise point positioning (PPP) with Galileo's High Accuracy Service (HAS), satellite-based augmentation system (SBAS) solutions and standalone operation. Three-unit phase comparisons separate component noise and characterize disciplined-output stability over 200\,000 s recordings. Across five configurations, each analysed over 200\,000 s, equivalent single-unit differential Allan deviation and time deviation at 1000 s span $3.5\times10^{-13}$--$1.0\times10^{-12}$ and 115--447 ps, respectively. The maximum equivalent time deviation over averaging times from 1 to 50\,000 s spans 143--1090 ps. These results support applications in astronomical timing, including low-frequency very long baseline interferometry (VLBI).
\end{abstract}

\section{Introduction}\label{sec:intro}
Accurate timing is essential for synchronizing astronomical instruments and studying rapidly varying phenomena. Applications include very long baseline interferometry (VLBI) \citep{Broten1967,MenaParra2022}, photon time tagging \citep{Lau2023SPINA}, optical intensity interferometry \citep{Cassanelli2025IQUEYE}, and fast-transient detection and analysis \citep{Lau2025PhotonDetection}. These applications combine stable local frequencies with consistent event epochs.

Global navigation satellite systems (GNSS), such as the Global Positioning System (GPS), provide position and time from satellite signals \citep{Milliken1978}. Precise point positioning (PPP) combines code and carrier measurements with precise satellite orbit and clock products \citep{Zumberge1997}; Galileo's High Accuracy Service (HAS) supplies correction products without a local reference station \citep{GalileoHAS2023}.

Here we present SimpleClock, a low-cost, high-stability timing reference that uses GNSS to discipline a local oscillator for astronomical instrumentation. Frequency and phase feedback can be enabled independently to suit the available GNSS solution. Short three-unit component studies guide the control settings; longer discipline-on recordings assess delivered-output stability. Although developed for astronomy, SimpleClock may also serve as a laboratory frequency reference and a timing source for distributed instruments.

\Needspace{7\baselineskip}
\section{Hardware and embedded implementation}\label{sec:hardware}
\begin{figure*}[t]
\centering
\includegraphics[width=0.96\textwidth]{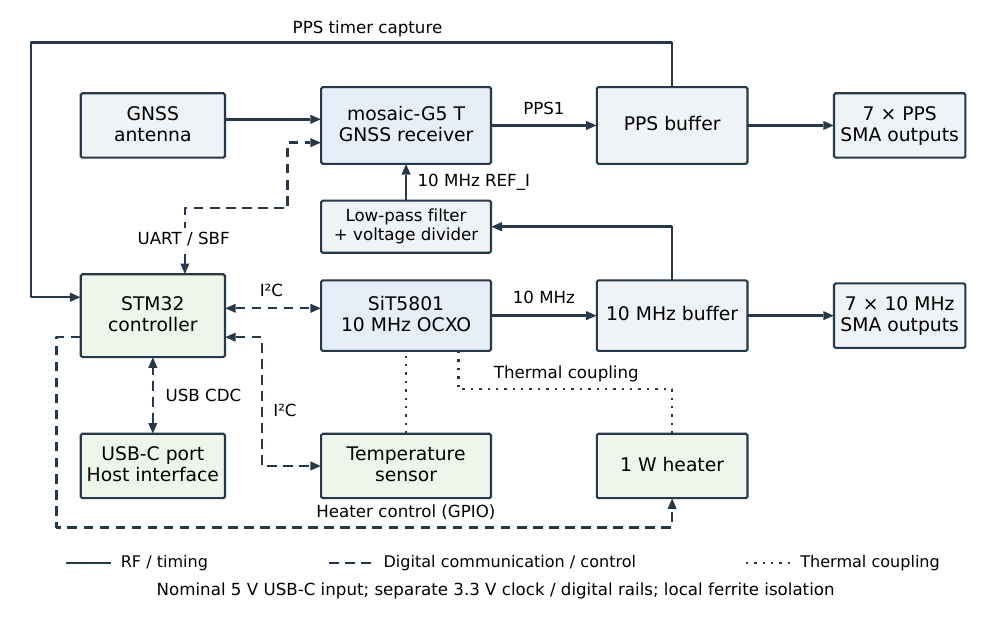}
\caption{SimpleClock architecture. Solid arrows denote radio-frequency (RF) and timing signals, dashed arrows digital communication and control, and dotted lines thermal coupling. The two I$^2$C links share a bus; USB-C connects to the microcontroller. \texttt{REF\_I} and \texttt{PPS1} label the receiver reference input and PPS output. PPS is derived coherently from the external 10 MHz reference.}\label{fig:architecture}
\end{figure*}

SimpleClock combines a GNSS timing receiver, a digitally tunable microelectromechanical systems (MEMS) oven-controlled oscillator (OCXO), external thermal regulation and a microcontroller (Fig.~\ref{fig:architecture}). The receiver measures clock bias and drift, and the microcontroller uses these observables to steer the oscillator while regulating the external heater.

\subsection{Reference and output paths}
A Septentrio mosaic-G5 T receiver supplies GNSS observables, and a 10 MHz SiTime SiT5801 MEMS OCXO supplies the physical frequency reference. The installed oscillator uses 3.3 V power, digital tuning over an inter-integrated circuit (I$^2$C) bus and a specified fractional tuning resolution of $5\times10^{-14}$ \citep{SiTime5801}. We call the applied fractional-frequency adjustment the oscillator trim.

The multiband GNSS antenna and coaxial cable are external to the receiver module (Fig.~\ref{fig:hardware-components}). A compatible active antenna must cover the navigation bands in use, present a 50 $\Omega$ interface and provide 15--50 dB net gain after cable loss in this single-antenna configuration \citep[Sec.~4.2]{SeptentrioHardware}. Satellite reception of HAS corrections additionally requires Galileo E6-B coverage at 1278.75 MHz \citep{GalileoHAS2023}.

Two one-to-eight buffers distribute seven subminiature version A (SMA) coaxial outputs each for 10 MHz and pulse-per-second (PPS) signals. The remaining frequency branch feeds the receiver external-reference input through a low-pass filter and voltage divider; the remaining PPS branch feeds the microcontroller. The receiver has the \texttt{FreqSync} capability enabled and uses the OCXO reference with internal continuous clock steering disabled. Its \texttt{RxClock} mode generates one PPS pulse per $10^7$ input cycles, maintaining coherence with the 10 MHz reference during GNSS outages \citep[Appendix~D]{SeptentrioHardware}. When feedback is enabled, the external controller disciplines the OCXO using receiver observables. Referencing these observables to the same OCXO removes the need for an external time-interval measurement circuit \citep[Appendix~D]{SeptentrioHardware}. User delay-compensation settings remain at their defaults for this differential-stability experiment.

Receiver commands and Septentrio Binary Format (SBF) records use a universal asynchronous receiver/transmitter (UART). The I$^2$C bus controls the oscillator and temperature sensor. Universal Serial Bus (USB) communications device class (CDC) carries configuration and diagnostics through a Type-C (USB-C) port. General-purpose input/output (GPIO) controls the heater.

\subsection{Power and thermal assembly}
The USB-C port supplies nominal 5 V input power. Separate 3.3 V regulator branches and ferrite isolation are intended to reduce shared supply impedance between digital, receiver and oscillator loads. An external 1 W thin-film heater and local temperature sensor stabilize the oscillator's surroundings, independently of its internal oven. Sensor lag and spatial gradients can affect measured temperature--frequency correlations.

\begin{figure}[tbp]
\centering
\begin{minipage}[b]{0.32\linewidth}
\centering
\includegraphics[width=\linewidth]{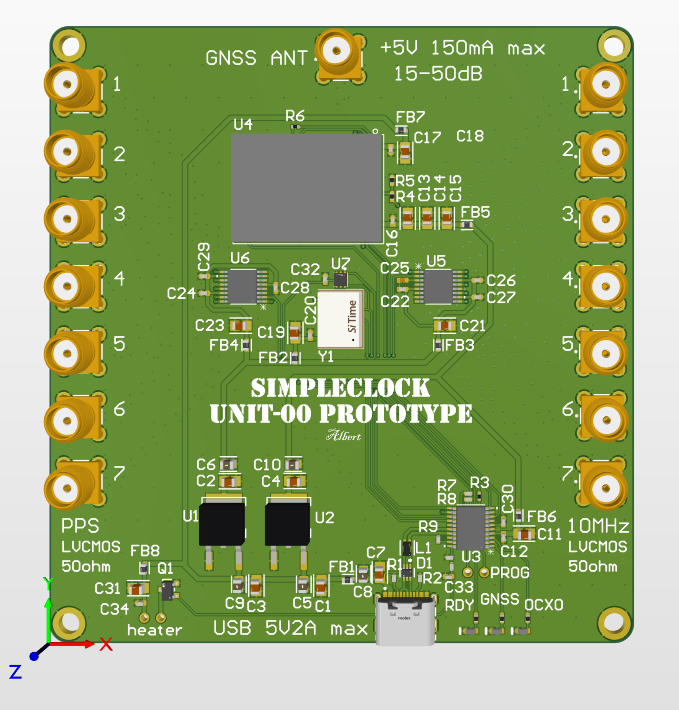}
\par\smallskip\small (a) PCB rendering
\end{minipage}\hfill
\begin{minipage}[b]{0.27\linewidth}
\centering
\includegraphics[width=\linewidth,trim=0bp 140bp 0bp 100bp,clip]{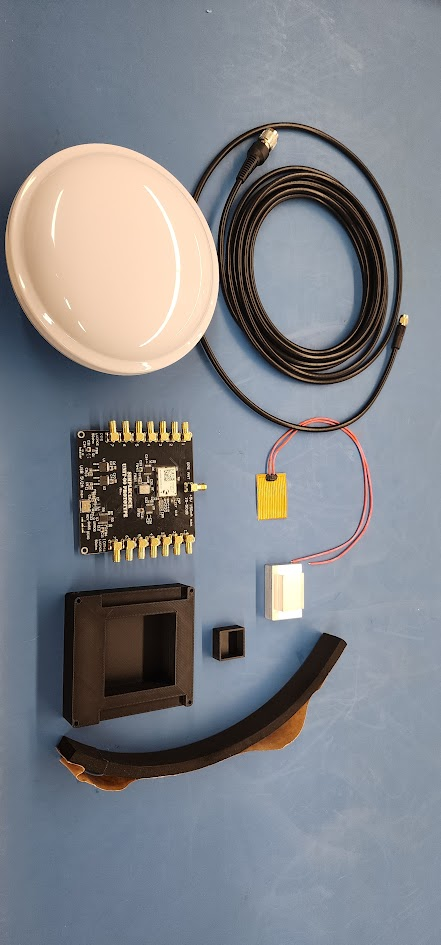}
\par\smallskip\small (b) Components
\end{minipage}\hfill
\begin{minipage}[b]{0.36\linewidth}
\centering
\includegraphics[trim=35bp 406bp 68bp 440bp,clip,angle=90,width=\linewidth]{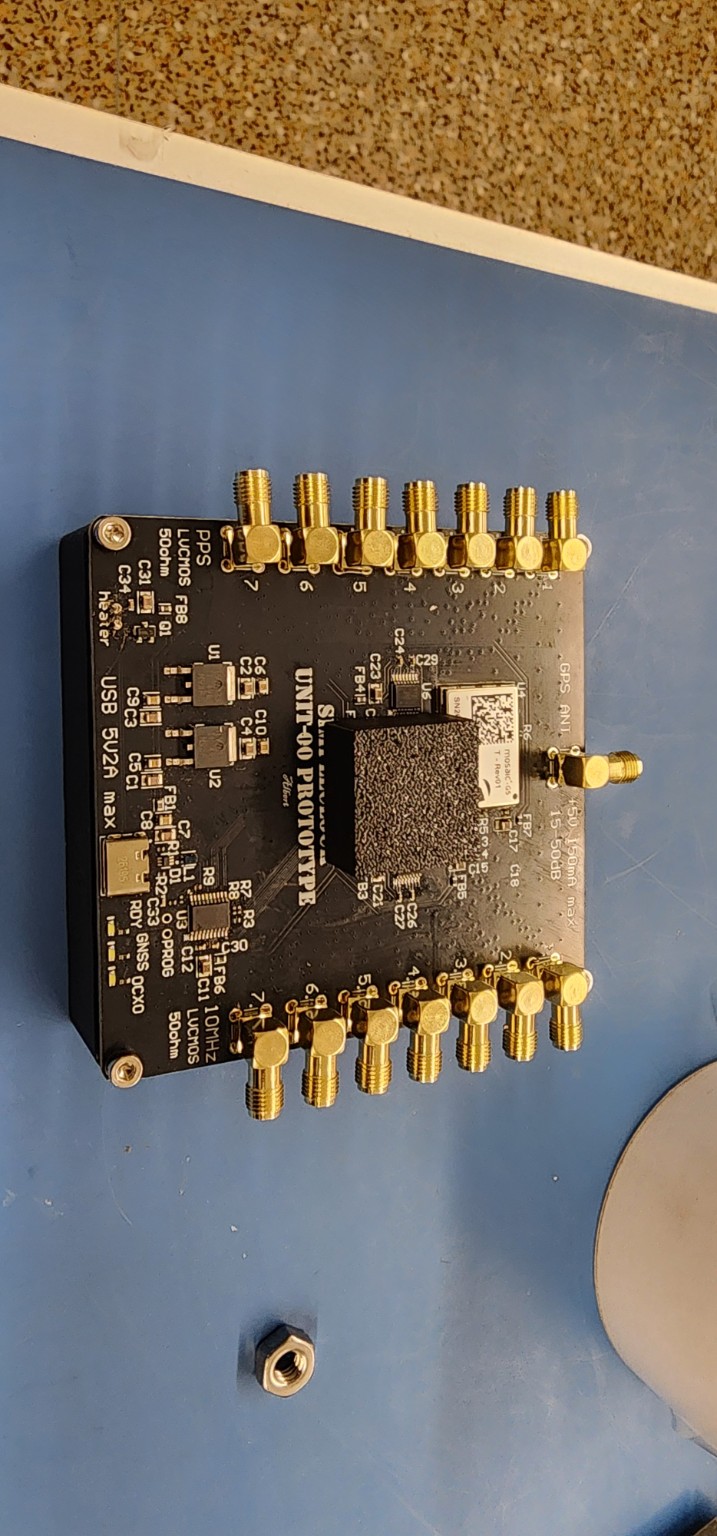}
\par\smallskip\small (c) Assembled module
\end{minipage}
\caption{SimpleClock hardware, with independent panel scales. (a) Computer-aided design rendering of the PCB. (b) Multiband GNSS antenna/cable (white, top), soldered PCB (middle left), 1 W heater (yellow), approximately 55 g aluminium block (below heater), lower/upper printed shields (lower left/centre), and insulation foam (bottom). (c) Assembled module: PPS outputs left, 10 MHz right, antenna connector top and USB-C bottom; heater and aluminium block are concealed.}\label{fig:hardware-components}\label{fig:photo}
\end{figure}

The thermal buffer is an approximately 55 g aluminium block sandwiched between the heater and printed circuit board (PCB). Double-sided adhesive tape bonds the heater directly to the block; a small thermally conductive tape pad couples the block to the PCB directly beneath the OCXO. The block can have any shape that maintains thermal contact and electrical clearance. Lower and upper three-dimensionally printed shields, insulation foam and foam double-sided tape provide thermal insulation around the assembly (Fig.~\ref{fig:hardware-components}).

\subsection{Firmware and operating states}
An STM32 microcontroller runs an autonomous event loop. Clock updates follow accepted 1 s receiver epochs; heater control uses fixed 1 s frames. The clock and thermal controllers operate independently: thermal status is diagnostic and does not gate clock acquisition or tracking. PPS capture associates receiver epochs with local events; USB carries asynchronous configuration and diagnostics. Framing, cyclic redundancy check (CRC), epoch and status checks reject invalid, stale or duplicate measurements before control updates.

All enabled feedback modes perform coarse frequency acquisition and verification before tracking, including phase-only operation. Tracking prefers usable PPP/HAS, then satellite-based augmentation system (SBAS) solutions such as the Wide Area Augmentation System (WAAS), then standalone. GNSS loss or receiver data older than 2.5 s holds the last verified trim and stops clock integration. Recovery checks residual drift and captures a fresh local phase reference; the thermal controller continues independently. Appendix~\ref{app:clock} gives the qualification and recovery rules.

\section{Controller architecture and parameter selection}\label{sec:control-design}
\subsection{Receiver timing observables}\label{sec:observables}
Code, carrier-phase and Doppler measurements are native GNSS observables \citep[Chap.~5]{HofmannWellenhof2008} and are generated independently of the receiver's positioning correction mode \citep[Sec.~2.2]{SeptentrioFirmware}.

The controller uses two timing observables from the GNSS receiver: clock bias and clock drift. The receiver reports \texttt{RxClkBias} in milliseconds and \texttt{RxClkDrift} in parts per million (ppm), positive when its clock leads the reported GNSS time system or runs faster, respectively \citep[pp.~233, 237]{SeptentrioFirmware}. Writing bias in nanoseconds as $b_{\rm ns}$ and drift in parts per billion (ppb) as $d_{\rm ppb}$, the conversions are
\begin{equation}
 b_{\rm ns}=10^6\,\texttt{RxClkBias},\qquad
 d_{\rm ppb}=10^3\,\texttt{RxClkDrift}.\label{eq:units}
\end{equation}
Below, $b$ denotes bias in seconds and $d$ dimensionless fractional drift. The public guide identifies pseudorange and Doppler as inputs to the navigation solution \citep[Sec.~2.4]{SeptentrioFirmware}. The discipline-off component studies compare bias and drift residuals separately, allowing for different filtering and effective gates (Appendix~\ref{app:components}, Figs.~\ref{fig:component-rover-sbas}--\ref{fig:campaign-mask}). These receiver-residual statistics are distinct from the measured disciplined-output stability in Sec.~\ref{sec:campaign-exports}.

With \texttt{PVTMode = Static}, the position, velocity and time (PVT) solution reports a constant position. This position constraint is separate from the \texttt{Motion = Static} option of \texttt{setReceiverDynamics}, which tells the navigation filter to expect a stationary antenna and adjusts its motion/noise model \citep[p.~97]{SeptentrioFirmware}. A stationary antenna can therefore still be operated in rover PVT mode, with its position estimated rather than held constant. The bias and drift used here are PVT-solution fields. The measurements use receiver firmware v1.1, which supports PPP/HAS. We cite the public v1.0.1 guide for shared interface and field definitions. Comparisons use consistent time systems, signs and delays, with discontinuities treated as separate segments.

\subsection{Frequency and phase feedback}
The frequency branch corrects the receiver-reported fractional-frequency offset, while the phase branch corrects accumulated clock bias. Either branch can operate alone; using both lets the frequency branch respond to short-term drift while the phase branch controls slower timing changes. Their relative gains are chosen from the component studies (Sec.~\ref{sec:control-thermal}).

Let $u$ be the dimensionless oscillator trim and $d$ the receiver-reported fractional drift. The phase error is $e=b-b_{\rm ref}$ (s), relative to a bias anchor $b_{\rm ref}$ captured when phase feedback starts. Proportional--integral (PI) feedback \citep[Chap.~10]{AstromMurray2008} produces the fractional-frequency command $w$:
\begin{align}
 \dot I_\phi&=-K_i e,\qquad w=I_\phi-K_p e,
 \label{eq:clock-cascade}
\end{align}
where $I_\phi$ is the integral contribution. The parameter $T_\phi$ (s) sets $K_p=2/T_\phi$ (s$^{-1}$) and $K_i=T_\phi^{-2}$ (s$^{-2}$).

In phase-only operation, $w$ directly sets the oscillator trim, $u=w$. With both branches enabled, $w$ instead sets a bounded drift target $v$ for the frequency-locked loop (FLL), whose nominal update is $\dot u=(v-d)/T_f$; $T_f$ (s) sets the frequency-loop gain. Frequency-only operation uses $v=0$, and disabling both branches holds the last trim. Thus the same phase controller either steers the oscillator directly or supplies the target for the frequency branch.

Bias feedback stabilizes timing around the local anchor; absolute pulse-per-second alignment with Coordinated Universal Time (UTC) additionally requires epoch and instrumental-delay calibration. The firmware preserves trim continuity when feedback branches or gains change. Appendix~\ref{app:clock} gives the discrete 1 Hz updates, input guards, actuator limits and transition rules.

\subsection{Experimental tuning and thermal regulation}\label{sec:control-thermal}
The discipline-off component studies (Appendix~\ref{app:components}) and separate short discipline-on tuning runs, not shown, guide the empirical feedback settings in Table~\ref{tab:campaign-runs}. The long-run campaign in Sec.~\ref{sec:campaign-exports} evaluates sustained output stability with these settings.

A parallel proportional--integral--derivative (PID) thermal controller uses filtered temperature for the proportional/integral terms, separately filtered derivative-on-measurement, bounded duty and anti-windup \citep[Chap.~10]{AstromMurray2008}. With approximately 6 K of heater authority, an outer integrator slowly moves the temperature goal to center actual duty near 50\% and preserve control headroom. GNSS disciplining is intended to track the associated slow frequency drift. Thermal regulation and goal adaptation operate independently of GNSS and clock holdover, subject to their own thermal tracking and safety conditions. Appendix~\ref{app:thermal} gives startup goal selection and model-guided thermal settings. Discipline-off clock tests hold trim while recording the thermal-controller state.

\section{Measurement campaign}\label{sec:results}
\subsection{Rooftop configurations and satellite selection}
\begin{figure}[!htbp]
\centering
\begin{minipage}[b]{0.31\linewidth}
\centering
\includegraphics[trim=0bp 421bp 0bp 250bp,clip,width=\linewidth]{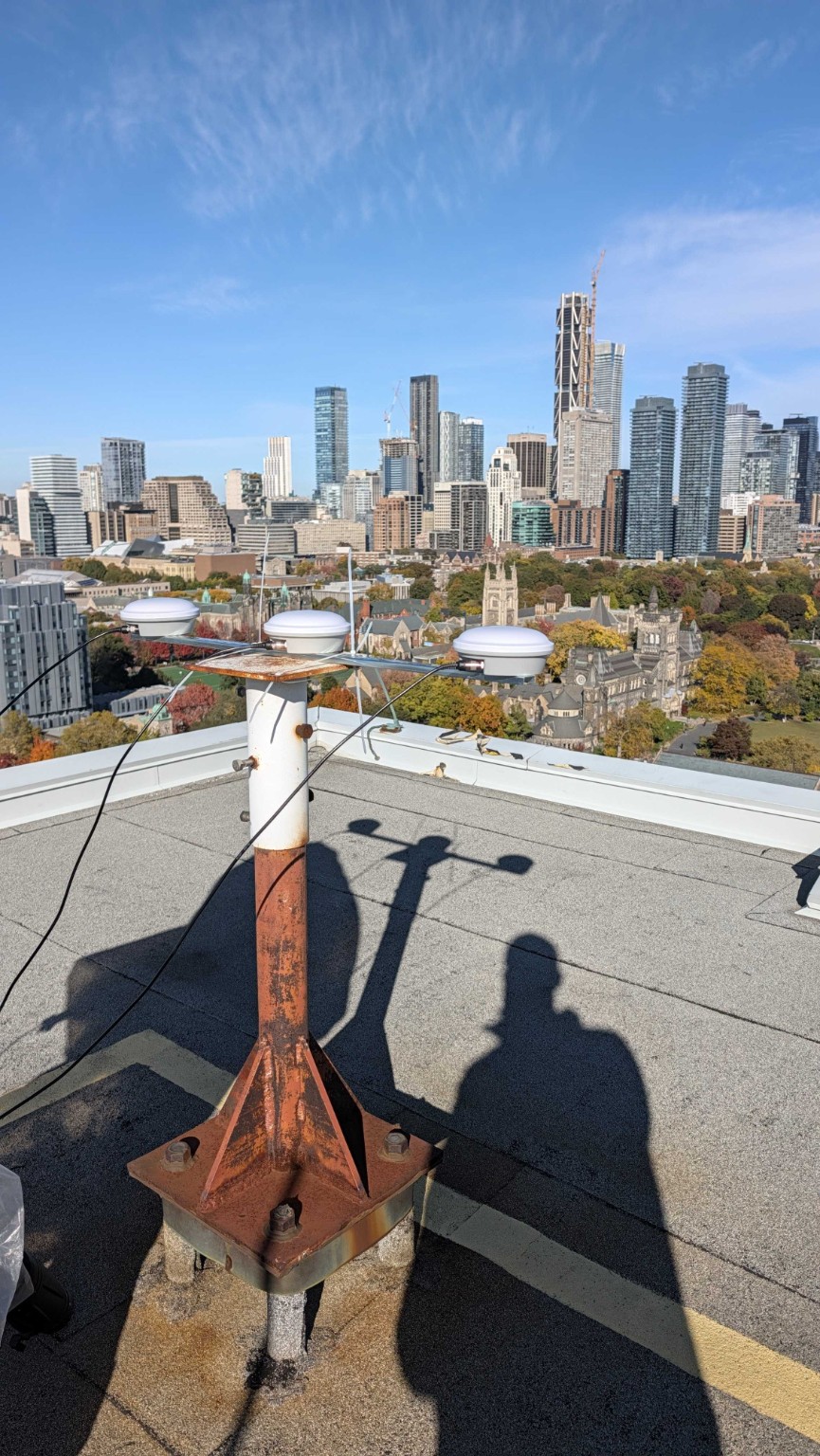}
\par\smallskip\small (a) Rooftop
\end{minipage}\hfill
\begin{minipage}[b]{0.665\linewidth}
\centering
\includegraphics[width=\linewidth]{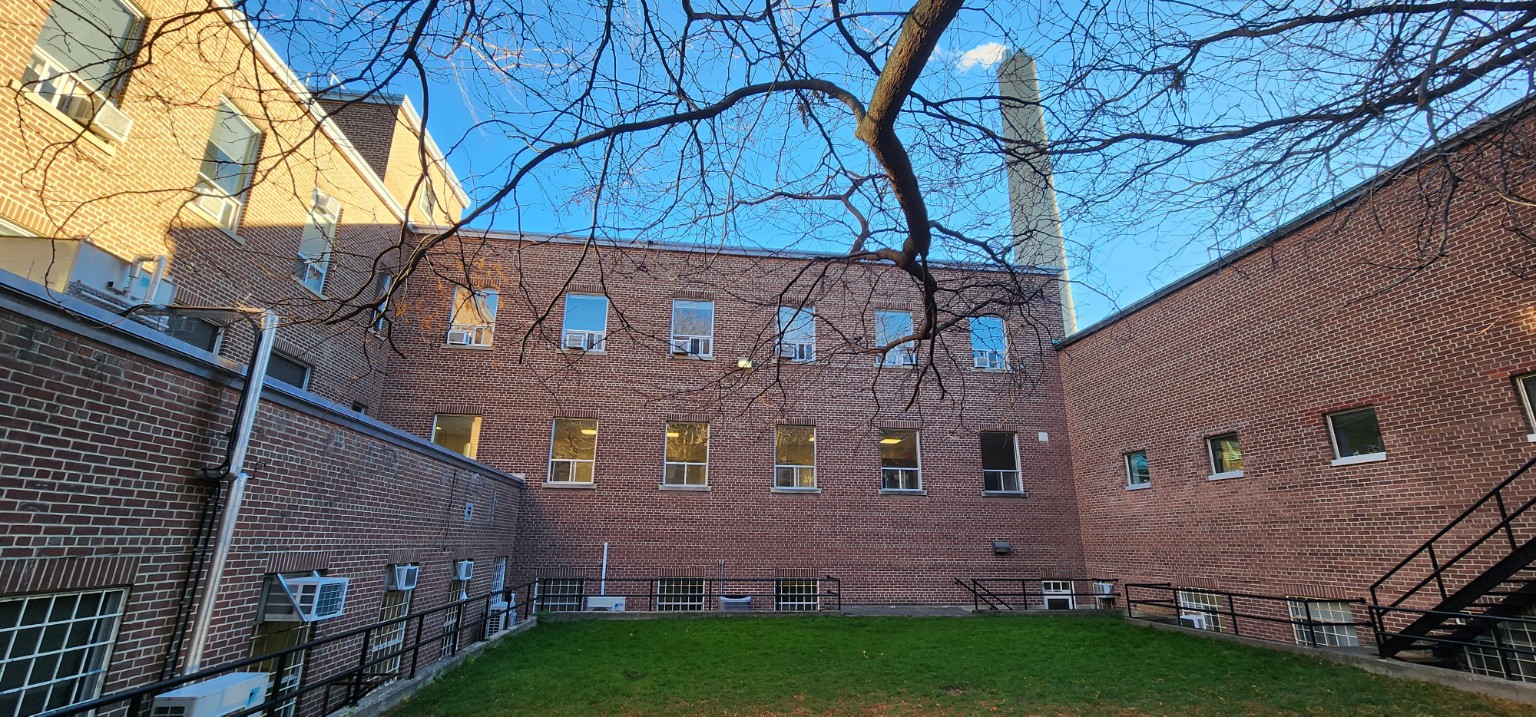}
\par\smallskip\small (b) Courtyard
\end{minipage}
\caption{GNSS test environments at the University of Toronto. (a) Three antennas mounted at the McLennan Physical Laboratories rooftop observatory. (b) The departmental courtyard, where nearby building walls and tree branches restrict sky visibility. The photographs illustrate the environments; the approximate visibility limits are described in the text.}\label{fig:test-sites}
\end{figure}

The measurements were conducted at two sites on the University of Toronto's St. George campus in Toronto, Canada. Initial tests used the courtyard of the David A. Dunlap Department of Astronomy and Astrophysics building (approximately $43.661^\circ$ N, $79.398^\circ$ W), referred to below as the confined yard (Fig.~\ref{fig:test-sites}). We then moved the antennas to the rooftop observatory of the McLennan Physical Laboratories building (approximately $43.661^\circ$ N, $79.399^\circ$ W) to improve sky visibility. The rooftop provides visibility down to approximately $10^\circ$ elevation above the horizon; obstructions at the courtyard extend to approximately $45^\circ$. We use the rooftop for the main stability campaign and retain the yard records as a separate test of performance under restricted visibility (Sec.~\ref{sec:confined-sky}).

We compare three units, A, B and C, in five rooftop configurations: standalone \texttt{PVTMode = Rover}, SBAS, masked PPP/HAS, standalone \texttt{PVTMode = Static}, and unmasked PPP/HAS. Each main analysis covers 200\,000 s (55.56 h), with the selected receiver solution maintained throughout; no fallback occurred during these intervals. Appendix~\ref{app:components} presents shorter discipline-off studies used to identify relative noise levels and crossover timescales. Section~\ref{sec:confined-sky} reports confined-sky recordings as a separate test of site sensitivity.

The configurations separate correction service from position constraint. Standalone operation uses broadcast navigation data without SBAS or HAS corrections; SBAS adds regional satellite-broadcast corrections, using WAAS in these tests. PPP/HAS uses precise orbit, clock and bias corrections supplied by Galileo HAS \citep{GalileoHAS2023}. With \texttt{PVTMode = Rover}, position is estimated as part of the navigation solution; \texttt{PVTMode = Static} holds position constant after its initialization (Sec.~\ref{sec:observables}). Unmasked PPP/HAS is the preferred open-sky configuration, while standalone \texttt{PVTMode = Static} is a practical choice for stationary instruments under restricted visibility. The masked runs diagnose the effects of satellite selection.

Navigation-satellite exclusions on B and C keep their used satellite sets below approximately 50\% overlap. We call this configuration masked; unmasked denotes operation without these imposed exclusions. A uses all eligible navigation satellites without imposed satellite exclusions in every configuration and provides the third comparison needed for the TCH estimates. Used navigation satellites, available satellites and correction broadcasters are distinct: tracking can continue for satellites excluded from the PVT solution \citep[pp.~75, 100]{SeptentrioFirmware}.

PPP/HAS typically used approximately 12--15 navigation satellites in these tests, compared with more than 30 in the other modes. Masking reduced this count to around eight and degraded performance, motivating the additional unmasked PPP/HAS run to assess the effect of satellite exclusions.

In the short PPP/HAS test, B and C each had 6--12 available navigation satellites with exclusions, compared with 12--20 without them. The selected interval keeps both counts at eight or more because lower counts produced instability in this setup. Using all otherwise eligible satellites offers greater redundancy and potentially better geometry than the masked 8--10-satellite configuration.

Correction inputs remain common despite PVT exclusions: HAS data from PVT-excluded broadcasters are applied, and the few visible WAAS broadcasters are common to both receivers. Sharing each service's products is expected, while residual correction errors may cancel in pair differences. Fewer usable satellites can increase individual receiver noise, whereas reduced overlap can expose previously cancelled common errors. Masked/unmasked comparisons therefore probe count, geometry and covariance together (Appendix~\ref{app:covariance}).

\subsection{Synchronized output-phase comparison}\label{sec:threeclock}
\begin{figure*}[t]
\centering
\includegraphics[width=0.88\textwidth]{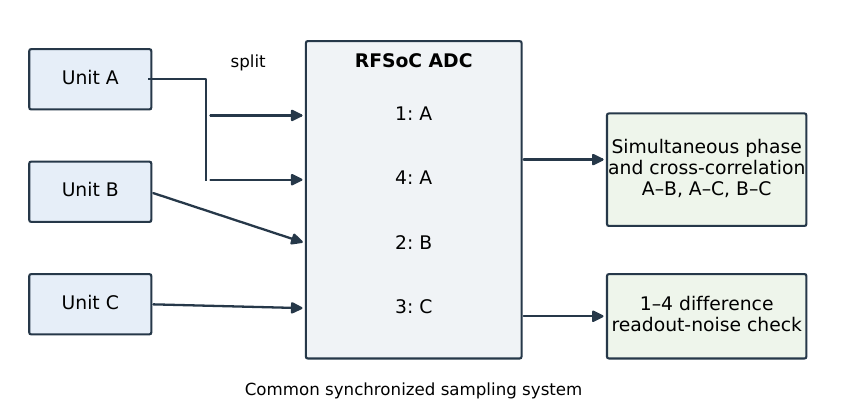}
\caption{Simultaneous comparison of A, B and C using radio-frequency system-on-chip (RFSoC) analog-to-digital converter (ADC) channels. Channels 1 and 4 share A to assess differential readout noise through cancellation of the common input phase. Receiver observables are aligned with the output-phase epochs.}\label{fig:threeclock}
\end{figure*}

Four synchronized analog-to-digital converter (ADC) channels on an RFSoC 4x2 radio-frequency system-on-chip (RFSoC) board digitize the 10 MHz OCXO outputs at $4.952\times10^9$ samples s$^{-1}$ per channel: unit A feeds channels 1 and 4 (ADC A and D), B feeds channel 2 and C feeds channel 3 (Fig.~\ref{fig:threeclock}). Firmware cross-correlates channel pairs at several harmonics; we analyse the 39th harmonic at 390 MHz. This practical choice uses existing analysis software and an odd harmonic with useful signal power, while providing greater phase sensitivity to time delay than the 10 MHz fundamental. Each phase estimate averages 100 ms of RFSoC data, with timing resolution of a few picoseconds. The duplicated A input measures differential splitter/cable/readout noise; common sampling phase cancels in the pair difference.

The phase difference at the selected tone gives pair time error $x_{ij}=\Delta\varphi_{ij}/(2\pi f_r)$, with $f_r=390$ MHz and pair ordering $x_{ij}=x_i-x_j$ \citep[Sec.~2]{Riley2008}. The single-tone phase has a time-wrapping period of $1/f_r\simeq2.56$ ns. The duplicated-input comparison has overlapping Allan deviation (ADEV), denoted $\sigma_y(\tau)$, of approximately $3\times10^{-13}$ at 1 s and $6\times10^{-16}$ at 1000 s, at least an order of magnitude below the clock-pair instability over the common plotted range of the short component studies. These measurements characterize the 10 MHz output; PPS additive jitter and absolute connector delays are not separately characterized. Appendix~\ref{app:readout} discusses the cancellation of common sampling error.

We compute overlapping ADEV, $\sigma_y(\tau)$, and time deviation (TDEV), $\sigma_x(\tau)$, directly from measured pair phase using standard estimators \citep[Secs.~5.2.4--5.2.6]{Riley2008}. Appendix~\ref{app:analysis} summarizes the processing and statistical conventions.

Each unit supplies receiver bias/drift and solution-quality information at 1 Hz over USB, together with controller and external thermal-controller state. The main campaign enables oscillator disciplining; the component studies in Appendix~\ref{app:components} hold the frequency word fixed with discipline disabled. Analysis aligns all streams on a 1 s grid and unwraps RFSoC phase within each continuous segment. No receiver, controller or RFSoC resets occurred during data capture. All three pair statistics within each data product use identical valid time windows. The selected receiver solution is maintained throughout each long-run analysis interval. The short component studies retain epochs only when all three receivers use the mode under study. The long-run output analysis retains the response to normal thermal regulation as part of delivered performance. The OCXO-only thermal exclusions and the two-unit exception in the short unmasked record are described in Appendix~\ref{app:components}.

\subsection{Three-cornered-hat estimates and covariance}
For independent sources with negligible or independently characterized readout error, the three-cornered-hat (TCH) method follows from addition of pair Allan variances \citep{Gray1974}:
\begin{equation}
 \widehat{\sigma}_A^2(\tau)=\frac{\sigma_{AB}^2(\tau)+\sigma_{AC}^2(\tau)-\sigma_{BC}^2(\tau)}{2}.\label{eq:tch}
\end{equation}
Cyclic expressions give B and C; variances are combined before taking square roots. The same operation applies to identically filtered TDEV variances under corresponding independence assumptions. The main results apply this decomposition to measured disciplined-output phase; the component studies apply it separately to each reconstructed product.

Measured pair curves provide the primary differential-stability result. Individual-clock estimates use assumptions about covariance and readout error to partition that variance \citep{Premoli1993}. Shared satellite, atmospheric and correction errors can cancel from all pairs; free-running oscillators can share environmental disturbances. Negative variance estimates produce gaps in deviation curves. Finite records and dependence between pair estimates set the statistical uncertainty of the TCH decomposition \citep{Ekstrom2006}. Appendix~\ref{app:covariance} gives the covariance model.

The measured inputs are the pair time-difference series $x_{ij}(t)$ in seconds. From each series we calculate a pair deviation $s_{ij}(\tau)$: this denotes ADEV, $\sigma_{y,ij}(\tau)$, or TDEV, $\sigma_{x,ij}(\tau)$, according to the plotted panel. The primary curve is $s_{\rm eq}(\tau)=s_{BC}(\tau)/\sqrt2$, reported as an equivalent single-unit differential instability. For two equally noisy, independent clocks with negligible readout error, this equals either clock's individual deviation. We retain this normalization as a reporting convention without assuming that the measured clocks are independent.

For common windows and filters, $s_{\rm eq}^2$ also equals the mean of the B/C TCH variance estimates. This is an algebraic identity: A cancels from their average, although it can redistribute the individual B/C estimates. Appendix~\ref{app:covariance} gives the relation and its covariance limitations. The plots quantify relative clock agreement; errors shared by all three units require a separate comparison against a traceable reference.

\section{Timing performance}\label{sec:campaign-exports}
The five rooftop configurations introduced in Sec.~\ref{sec:results} are compared using the measured disciplined outputs. We then examine the confined-yard receiver records to assess sensitivity to restricted sky visibility.

\subsection{Long-duration disciplined outputs}
The long-run campaign covers the five configurations in Table~\ref{tab:campaign-runs}, with oscillator disciplining enabled throughout each accepted tracking interval. All five figures show measured B--C/$\sqrt2$ equivalent single-unit differential stability and the conditional B/C TCH estimates on the same 67 averaging times from 1 to 50\,000 s, including exactly 1000 s. We summarize ADEV and TDEV at $\tau=1000$ s and the maximum equivalent TDEV over $1\leq\tau\leq50\,000$ s; PPP/HAS maxima are evaluated on the plotted averaging-time grid.

\begin{table}[H]
\centering
\caption{Feedback settings and start times for the discipline-on datasets. Each analysed interval is 200\,000 s. Start times refer to the analysed intervals in UTC. Configurations follow figure order. Both denotes frequency and phase feedback; a zero $T_f$ disables frequency feedback.}\label{tab:campaign-runs}
\small
\begin{tabular}{@{}llrrl@{}}
\toprule
Configuration & Feedback & $T_f$ (s) & $T_\phi$ (s) & \shortstack[l]{Start time\\(UTC)} \\
\midrule
Standalone rover, masked & Both & 10 & 300 & \shortstack[l]{2026-10-29\\21:11:14} \\
SBAS, masked & Both & 10 & 300 & \shortstack[l]{2026-11-01\\04:58:40} \\
PPP/HAS, masked & Phase only & 0 & 10 & \shortstack[l]{2026-11-06\\03:35} \\
\shortstack[l]{Standalone, \texttt{PVTMode = Static}\\masked} & Both & 10 & 100 & \shortstack[l]{2026-10-27\\12:46:51} \\
PPP/HAS, unmasked & Phase only & 0 & 10 & \shortstack[l]{2026-11-03\\15:03:13} \\
\bottomrule
\end{tabular}
\end{table}

These empirical settings are guided by the discipline-off component studies in Appendix~\ref{app:components} and separate short discipline-on tuning runs, not shown. Combined feedback uses $T_f=10$ s, with $T_\phi=300$ s for standalone rover and SBAS and $T_\phi=100$ s for standalone \texttt{PVTMode = Static}. PPP/HAS uses phase-only feedback with $T_\phi=10$ s. The long recordings evaluate sustained output stability over thermal and satellite-geometry changes.

All five records are analysed over their final 200\,000 s, with 200\,001 consecutive phase samples at 1 s spacing and identical windows across all three pair series. No detrending, smoothing or outlier rejection is applied. For both PPP/HAS configurations, the receiver solution was allowed to settle for at least 2 h before data collection.

\begin{figure}[!htbp]
\centering
\includegraphics[width=\linewidth]{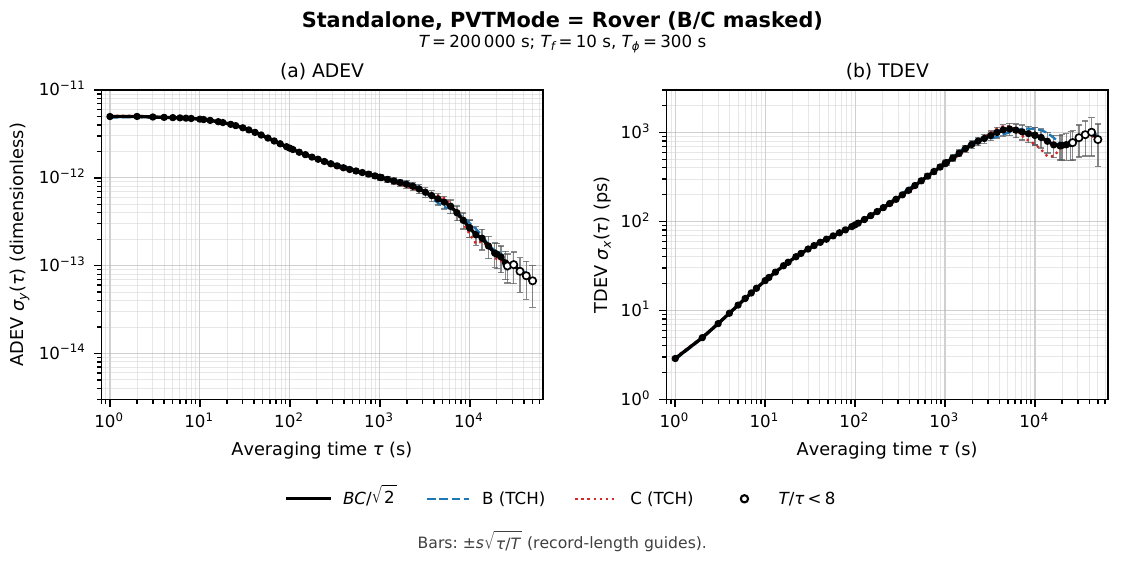}
\caption{Measured disciplined output in standalone \texttt{PVTMode = Rover} with B/C satellite exclusions. Loop settings are listed in Table~\ref{tab:campaign-runs}; plotting conventions are defined in Sec.~\ref{sec:campaign-exports}.}\label{fig:long-rover}
\end{figure}

\begin{figure}[!htbp]
\centering
\includegraphics[width=\linewidth]{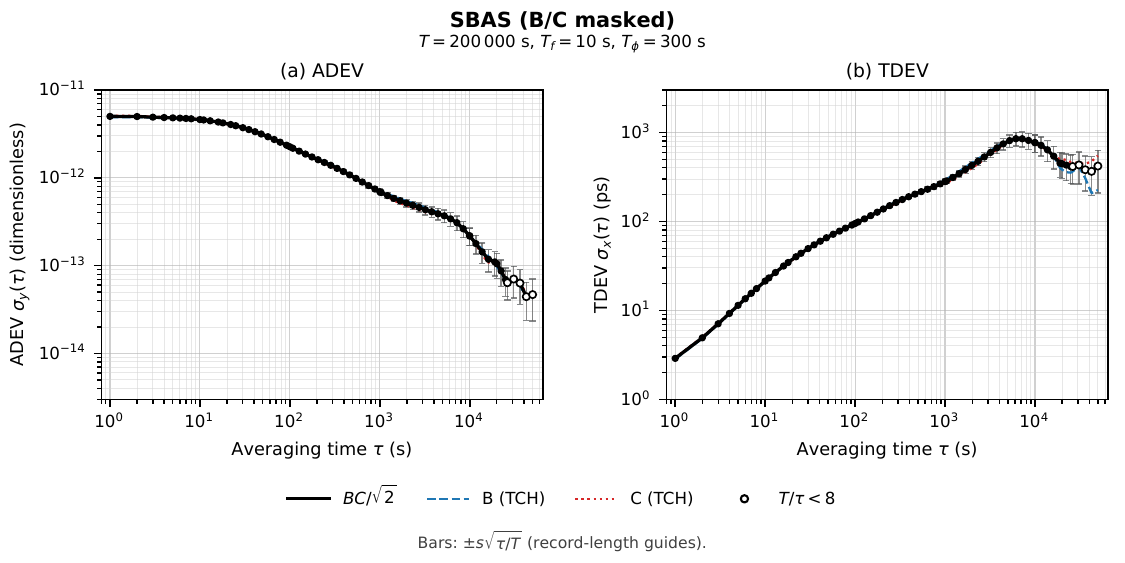}
\caption{Measured disciplined output in SBAS mode with B/C satellite exclusions.}\label{fig:long-sbas}
\end{figure}

\begin{figure}[!htbp]
\centering
\includegraphics[width=\linewidth]{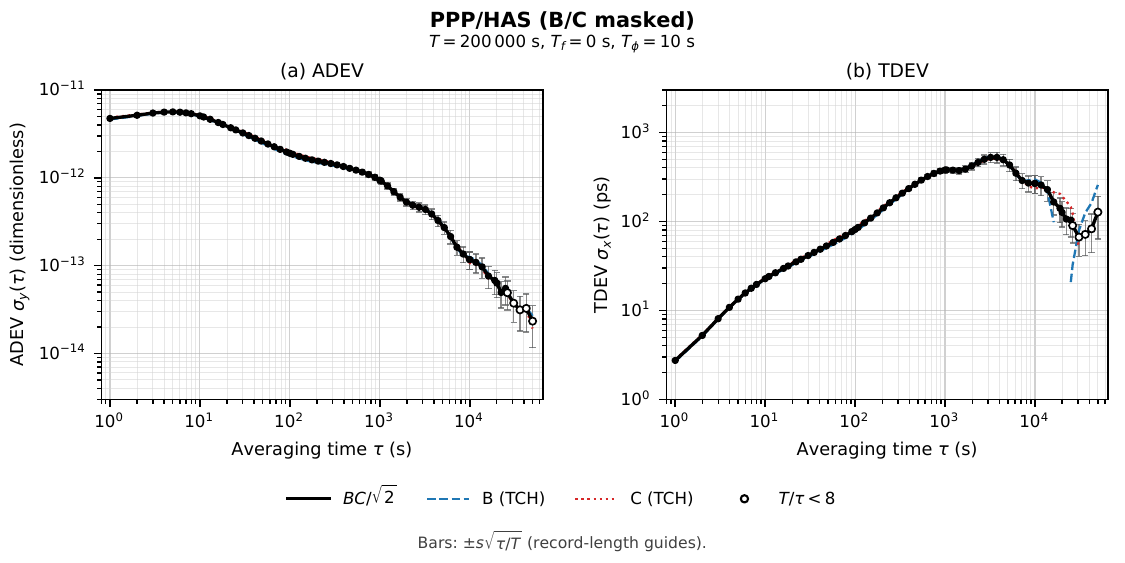}
\caption{Measured disciplined output in PPP/HAS mode with B/C satellite exclusions.}\label{fig:long-ppp-masked}
\end{figure}

\begin{figure}[!htbp]
\centering
\includegraphics[width=\linewidth]{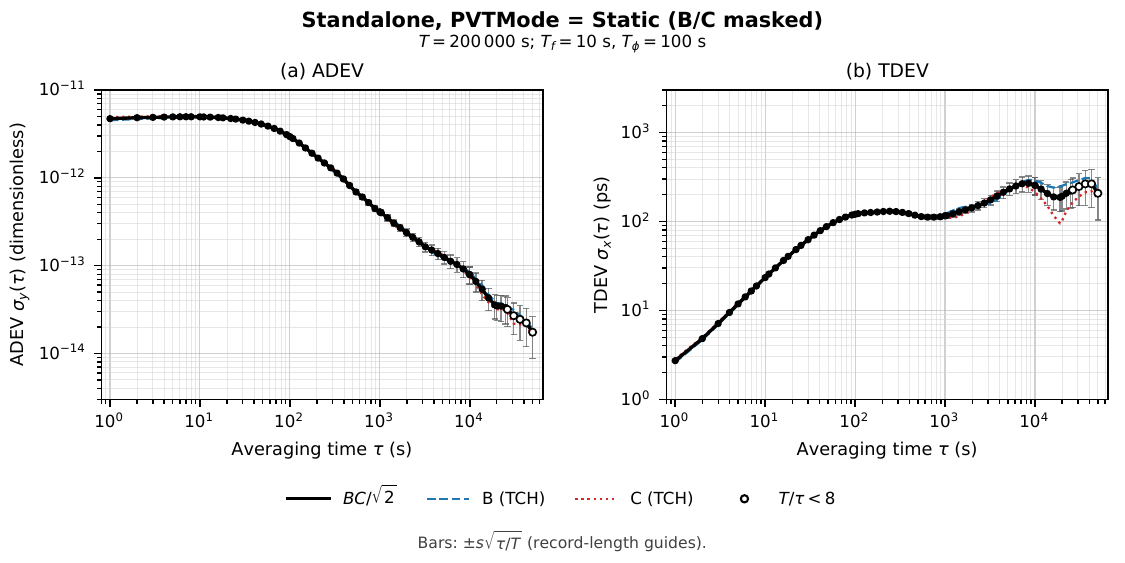}
\caption{Measured disciplined output in standalone \texttt{PVTMode = Static} with B/C satellite exclusions.}\label{fig:long-fixed}
\end{figure}

\begin{figure}[!htbp]
\centering
\includegraphics[width=\linewidth]{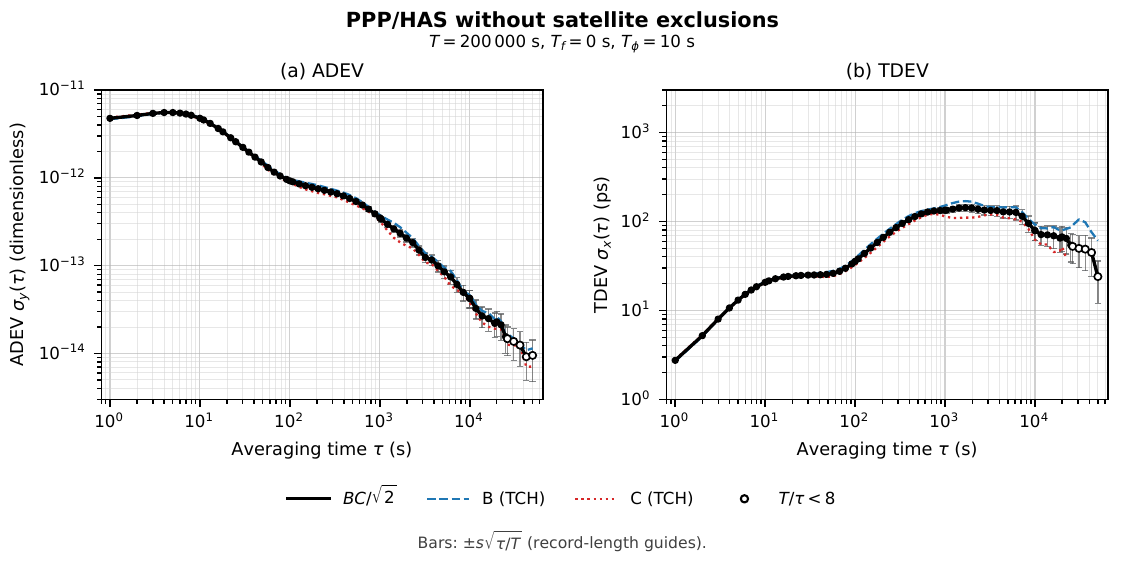}
\caption{Measured disciplined output in PPP/HAS mode without imposed satellite exclusions.}\label{fig:long-ppp-unmasked}
\end{figure}

Figures~\ref{fig:long-rover}--\ref{fig:long-ppp-unmasked} use common plotting conventions: ADEV at left and TDEV at right; black denotes B--C/$\sqrt2$, blue dashed and red dotted curves the B/C TCH estimates. Negative TCH variances give gaps. Error bars on the primary curve remain uncalibrated record-length guides, $\pm s\sqrt{\tau/T}$; hollow markers identify $T/\tau<8$. Appendix~\ref{app:covariance} gives the covariance interpretation. All records retain normal thermal regulation, with no detrending, smoothing or outlier rejection.

\begin{table}[H]
\centering
\caption{Equivalent single-unit differential stability, B--C/$\sqrt2$. ADEV is dimensionless; TDEV values are in picoseconds. Maxima cover $1\leq\tau\leq50\,000$ s, rounded to the closest 100s.}\label{tab:stability-summary}
\small
\begin{tabular}{@{}lrrrr@{}}
\toprule
Configuration & \shortstack{ADEV\\at 1000 s} & \shortstack{TDEV\\at 1000 s} & \shortstack{Maximum\\TDEV} & \shortstack{$\tau$ at\\maximum (s)} \\
\midrule
Standalone rover, masked & $1.0\times10^{-12}$ & 447 & 1090 & $5200$ \\
SBAS, masked & $6.9\times10^{-13}$ & 280 & 855 & $6600$ \\
PPP/HAS, masked & $9.4\times10^{-13}$ & 379 & 525 & $3200$ \\
Standalone static, masked & $4.1\times10^{-13}$ & 115 & 268 & $8500$ \\
PPP/HAS, unmasked & $3.5\times10^{-13}$ & 134 & 143 & $1700$ \\
\bottomrule
\end{tabular}
\end{table}

Table~\ref{tab:stability-summary} summarizes the five records. With the same loop settings, SBAS has 32\% lower ADEV and 37\% lower TDEV at 1000 s than standalone rover. Static standalone has approximately 2.5 and 3.9 times smaller values than rover, respectively, including the effect of its shorter phase-loop setting. Removing the PPP/HAS satellite masks gives 63\% lower ADEV and 65\% lower TDEV at 1000 s. These comparisons characterize complete configurations under different recording conditions; satellite exclusions change count, geometry and covariance together.

\FloatBarrier

\subsection{Performance under restricted sky visibility}\label{sec:confined-sky}
\begin{figure}[!htbp]
\centering
\includegraphics[width=\linewidth]{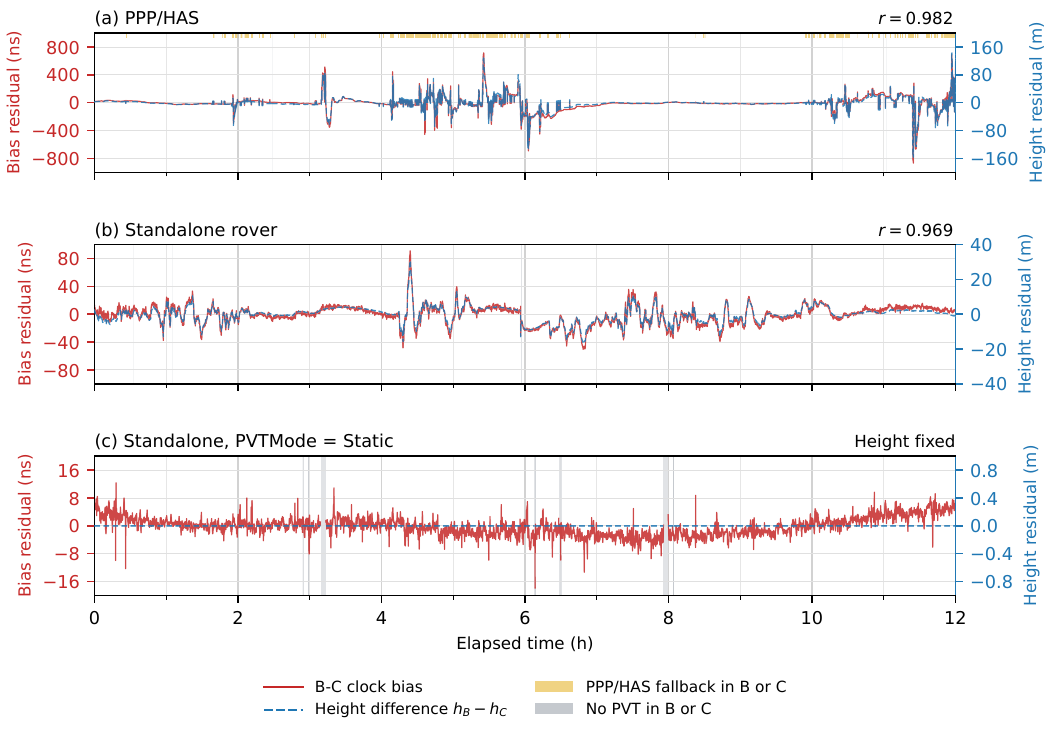}
\caption{Confined-sky B--C receiver clock bias (red, left axis) and height difference $h_B-h_C$ (blue, right axis), over the first 12 h of each separate record. Independent linear trends, including intercepts, are removed from both series on common valid samples; no smoothing or outlier rejection is applied. Vertical scales differ between panels and observables. Yellow marks PPP/HAS fallbacks and gray marks no-PVT intervals in B or C; gaps are retained. Height in \texttt{PVTMode = Static} is constrained, so its correlation is undefined. The PPP/HAS and rover records begin on 18 October 2026, and the \texttt{PVTMode = Static} record on 20 October.}\label{fig:confined-bias}
\end{figure}

The confined yard introduced in Sec.~\ref{sec:results}, with obstructions extending to approximately $45^\circ$ elevation, provides a test of site-specific robustness (Fig.~\ref{fig:confined-bias}). After linear-trend removal, B--C receiver-bias differences reach approximately 0.9 $\mu$s in the PPP/HAS configuration and 90 ns in standalone rover operation. Their correlations with the corresponding height-difference residuals are $r=0.982$ and 0.969, respectively. Standalone with \texttt{PVTMode = Static} has smaller detrended bias differences, within approximately 18 ns in the selected window, but still includes PVT outages. Its constant reported height is imposed by the positioning constraint.

The PPP/HAS record includes fallback solutions; no dedicated confined-sky SBAS comparison is available. These receiver-bias records characterize site sensitivity and position--clock coupling. The markedly smaller excursions demonstrate the robustness of \texttt{PVTMode = Static} in this restricted-sky environment, where multipath and interference are practical concerns; the open-rooftop discipline-on record separately characterizes its delivered-output stability.

\FloatBarrier

\Needspace{10\baselineskip}
\section{Discussion}\label{sec:discussion}
Astronomical timing requirements depend on the measured observable and the interval over which clock errors remain uncorrected. We consider radio interferometry and photon time tagging as representative applications of SimpleClock.

\subsection{Timing requirements for VLBI}\label{sec:astronomical-requirements}
VLBI requires stable relative station timing both during coherent averaging and between calibration observations \citep{Rogers1981}. These intervals constrain different aspects of performance: rapid timing fluctuations reduce coherence, while unmodelled delay evolution between calibrators can bias localization.

Let $\delta t(t)$ be the differential station timing residual after the intended delay and phase calibration, and $T_{\rm int}$ the interval over which complex visibilities are averaged without further phase correction. At observing frequency $\nu$, the retained amplitude fraction is
\begin{equation}
 C(T_{\rm int})=\left|\frac{1}{T_{\rm int}}\int_0^{T_{\rm int}}
 e^{i2\pi\nu\delta t(t)}\,dt\right|.
 \label{eq:coherence-interval}
\end{equation}
For zero-mean Gaussian residual phase, the ensemble complex-average amplitude is $e^{-(2\pi\nu\sigma_{\rm res})^2/2}$, where $\sigma_{\rm res}$ is the root-mean-square (rms) varying timing residual for that averaging and calibration scheme \citep{Rogers1981}. A nominal 1\% amplitude-loss budget therefore gives
\begin{equation}
 \sigma_{\rm res}\lesssim\frac{\sqrt{-2\ln(0.99)}}{2\pi\nu},\label{eq:coherence-budget}
\end{equation}
approximately 75 ps at 300 MHz and 45 ps at 500 MHz. These illustrate the 1\% coherence budget across the band of the Canadian-Chilean Array for Radio Transient Studies (CHARTS) \citep{Cassanelli2025CHARTS,Lau2026CHARTS}, a potential application of SimpleClock.

The measured unmasked PPP/HAS outputs retain at least 99\% mean coherence for integration times up to 41 s at 500 MHz and 156 s at 300 MHz; the corresponding standalone \texttt{PVTMode = Static} intervals are 17 s and 33 s. We evaluate Eq.~\ref{eq:coherence-interval} directly from each unscaled B--C phase record, linearly interpolating between its 1 s samples and averaging the coherence magnitude over all overlapping window starts in the 200\,000 s record. No frequency trend is removed. At 500 MHz, the mean amplitude loss for 10 s integrations is 0.25\% for unmasked PPP/HAS and 0.42\% for standalone \texttt{PVTMode = Static}. These results quantify the clock contribution directly, without converting ADEV or TDEV into a residual timing error.

As an external localization benchmark, the Canadian Hydrogen Intensity Mapping Experiment (CHIME) fast radio burst (FRB) Outriggers allocated approximately 200 ps to relative clock-delay error. Its interpolation model, with calibrators separated by about 2000 s, motivates $\sigma_y(1000\,\mathrm{s})\lesssim2\times10^{-13}$ \citep[Sec.~3 and Appendix~A]{MenaParra2022}. SimpleClock's measured B--C pair ADEV at 1000 s is $4.9\times10^{-13}$ for unmasked PPP/HAS and $5.8\times10^{-13}$ for standalone \texttt{PVTMode = Static}, the same order as this reference level (within factors of approximately 2.5 and 3, respectively). Wide-field arrays may permit more frequent or simultaneous calibrator observations and hence shorter intervals of unmeasured clock evolution \citep{Lanman2024KKO}, subject to correlated flux, sensitivity and direction-dependent propagation. This can relax stability requirements between calibrations without changing a chosen coherence-loss budget. Calibration cadence and GNSS-outage holdover remain separate design constraints.

The main deployment comparison is between unmasked PPP/HAS under open sky and standalone \texttt{PVTMode = Static} for restricted-sky sites (Sec.~\ref{sec:confined-sky}). Their relative performance depends on timescale and application: static standalone has slightly lower TDEV at 1000 s, while unmasked PPP/HAS provides longer mean coherence intervals and lower maximum TDEV over the evaluated range (Table~\ref{tab:stability-summary}). These results support selecting the receiver mode and feedback settings for the site's visibility and the required averaging interval.

\subsection{Photon time tagging}
Astronomical photon-counting instruments have demonstrated approximately 50--100 ps relative event-timing precision \citep{Barbieri2009Aqueye,Zampieri2021Vega}. To preserve this precision across independently timed instruments or long recordings, clock-induced timing variation should fit within the same error budget, preferably below the detector and time-tagger contribution. The relevant interval is the time between effective clock corrections or re-anchoring; for a constant uncorrected fractional-frequency error $y_0$, accumulated time error is $y_0T_{\rm tag}$ over an interval $T_{\rm tag}$. SimpleClock's 10 MHz and PPS outputs offer a practical reference upgrade for observatory instruments with compatible external-reference inputs. Achieving comparable absolute UTC accuracy also requires epoch and instrumental-delay calibration, distinct from the relative stability measured here.

\subsection{Scope of the co-located comparison}\label{sec:common-gnss-errors}
The satellite masks reduce B/C navigation-satellite overlap, but all three antennas observe the same local sky. Residual ionospheric and tropospheric delays, satellite orbit/clock errors and shared correction-product errors can therefore remain correlated and partly cancel from the pair differences. The TCH decomposition does not recover errors common to all units (Appendix~\ref{app:covariance}).

Published studies illustrate relevant scales: higher-order ionospheric terms changed PPP receiver-clock solutions by several picoseconds and some long time-link solutions by more than 25 ps \citep{Yang2020Ionosphere}; a HAS GPS/Galileo comparison reported 0.11--0.21 ns standard deviations against post-processed reference solutions on 7200--12\,295 km links \citep{Zhang2024HASTime}. These are results for different receivers, processing and conditions, rather than a measured floor or correction for SimpleClock. Our results establish co-located differential stability; geographically separated measurements and an independent reference would quantify the additional propagation and GNSS-system errors relevant to remote observatories.

\subsection{Cost and accessibility}
SimpleClock's estimated prototype hardware cost is approximately US\$500 per unit. The SRS FS725 rubidium frequency standard starts at US\$3995, about eight times higher \citep{SRSFS7252026}, before adding an external GNSS reference for disciplining. The Spectrum Instruments TM-4D GPS-disciplined reference used to clock the CHIME digitizer is one example of an astronomical GNSS reference \citep{MenaParra2022}. The Makerfabs implementation of the open-source Open Compute Project (OCP) Time Card, assembled with a u-blox NEO-M9N GNSS receiver and an OCXO, is listed at US\$1500 \citep{MakerfabsTimeCard2026}. These prices illustrate SimpleClock's accessibility alongside its demonstrated differential stability. The systems offer different interfaces and holdover capabilities, with rubidium-based systems offering potential for longer GNSS-outage holdover \citep{Lombardi2021Holdover}; quantitative stability comparisons require measurements under comparable conditions.

\section{Conclusions}
SimpleClock provides a GNSS-disciplined 10 MHz and PPS reference with autonomous thermal regulation and independently enabled frequency and phase feedback, at an estimated prototype hardware cost of approximately US\$500. Component measurements guide loop selection, while the long-run campaign evaluates delivered-output stability.

Across five 200\,000 s recordings, equivalent single-unit differential ADEV and TDEV at 1000 s span $3.5\times10^{-13}$--$1.0\times10^{-12}$ and 115--447 ps, respectively; maximum equivalent TDEV over 1--50\,000 s spans 143--1090 ps. Unmasked PPP/HAS and standalone \texttt{PVTMode = Static} retain 99\% mean clock-only coherence at 500 MHz for integrations up to 41 s and 17 s, respectively, supporting applications in low-frequency VLBI.

The confined-sky receiver-bias comparison favours \texttt{PVTMode = Static} under restricted visibility, complementing unmasked PPP/HAS as the preferred open-sky configuration. Together, these modes address observatories with different antenna-site conditions.

A future implementation could use SimpleClock to discipline a tunable atomic frequency standard and extend usable holdover during GNSS outages. GNSS would supply the long-term reference, while the atomic oscillator would maintain timing when GNSS is unavailable \citep{Lombardi2021Holdover}.

\section*{Acknowledgments}
We thank the Dunlap Institute for Astronomy \& Astrophysics for funding and continued support. The Dunlap Institute is funded through an endowment established by the David Dunlap family and the University of Toronto. We also thank Mike Williams, Telescope Demonstrator in the David A. Dunlap Department of Astronomy and Astrophysics, for his assistance in accessing the rooftop observatory site.

\section*{Data and software availability}
The SimpleClock project repository is \url{https://github.com/awklau/SimpleClock}. Raw measurement recordings and simulation time series are available from the authors on request owing to their size.

\section*{Use of generative artificial intelligence}
ChatGPT and Codex (OpenAI) and Claude (Anthropic) were used for design discussions, component selection, electronic-design library generation, firmware and analysis-code development, figure preparation, and manuscript drafting and editing. The authors directed the work and are responsible for its validation and final content.

\FloatBarrier
\renewcommand{\theHequation}{\Alph{appendixc}.\arabic{equation}}
\renewcommand{\theHfigure}{\Alph{appendixc}.\arabic{figure}}
\renewcommand{\theHtable}{\Alph{appendixc}.\arabic{table}}
\appendix{Clock-control implementation and tuning}\label{app:clock}
The nominal tracking updates below are applied at 1 Hz after receiver qualification and coarse frequency acquisition. Clock and thermal qualification are independent.

\subappendix{Qualified epochs and discrete actuator paths}\label{app:clock-discrete}
Cold startup requires a usable GNSS solution for more than 60 s. All enabled modes then perform coarse acquisition using a 50-sample drift average, followed by a 50-sample verification with a residual-mean threshold of 0.05 ppb. Acquisition corrections have a 0.05 ppb/s slew limit. Phase activation has a separate 120 s qualification interval. Thermal stability does not gate these stages.

The phase gains are coupled: $K_p=2/T_\phi$ and $K_i=T_\phi^{-2}$ for positive $T_\phi$. This parameter specifies the gains; the enabled loop configuration determines the resulting response.

Input guards clip bias innovations to $\pm2$ ns and bound frequency innovations to 0.25 ppb in magnitude. The controller applies no measurement low-pass filter. The equations below describe the nominal tracking regime between guard and actuator-limit interventions.

Bias is referred to a captured local anchor, $e_k=b_k-b_{\rm ref}$. Clock integration uses $\Delta t=1$ s per accepted contiguous receiver epoch, rather than elapsed microcontroller time. Missing, invalid or duplicate epochs do not cause catch-up integration. With an enabled, qualified phase branch,
\begin{align}
 I_{\phi,k+1}^{\rm trial}&=I_{\phi,k}-K_i\Delta t\,e_k,\nonumber\\
 w_{k+1}&=I_{\phi,k+1}^{\rm trial}-K_p e_k,
 \label{eq:discrete-control}
\end{align}
Cold/recovery phase entry preserves the actuator state for that epoch; tracking updates begin on subsequent qualified epochs. The state labels are BOTH (both branches), PHASE (phase only), FREQUENCY (frequency only) and OFF (neither branch). In dimensionless fractional-frequency units, the actuator trial is
\begin{equation}
 u_{k+1}^{\rm trial}=\begin{cases}
 u_k+A_k\dfrac{\Delta t}{T_f}(v_{k+1}-d_k),&\text{BOTH},\\[1mm]
 w_{k+1},&\text{PHASE},\\[1mm]
 u_k-A_k\dfrac{\Delta t}{T_f}d_k,&\text{FREQUENCY},\\[1mm]
 u_k,&\text{OFF},
 \end{cases}\qquad
 v_{k+1}=\operatorname{clip}(w_{k+1},-v_{\max},v_{\max}),
 \label{eq:branch-actuator}
\end{equation}
with $v_{\max}=0.1$ ppb $=10^{-10}$. PHASE interprets the integral as total trim; BOTH interprets it as a drift request. The firmware assumes unit actuator gain and uses the linear frequency-loop coefficient $\Delta t/T_f$. Frequency authority $A_k\in[0,1]$ ramps over 60 s of microcontroller time at entry/re-entry. Before phase qualification, BOTH retains its entry drift request. PHASE uses the same coarse acquisition and verification as the other enabled modes; OFF holds trim.

A zero time parameter disables its branch: positive $(T_f,T_\phi)$ selects BOTH, $(0,T_\phi)$ PHASE, $(T_f,0)$ FREQUENCY and $(0,0)$ OFF. Only enabled branches are evaluated. The selected discipline-on settings are listed in Table~\ref{tab:campaign-runs}; Appendix~\ref{app:component-curves} distinguishes the earlier exploratory simulations.

Qualified tracking uses a 0.003 ppb/s slew limit and a $\pm1$ ppb span about the acquired trim, within a $\pm10$ ppb envelope about the boot readback. These actuator bounds apply to phase-only as well as combined tracking. Conditional integration suppresses outward windup, including downstream limiting in BOTH, while allowing unwinding. Fractional trim is retained through word quantization and writes require matching independent readback. State preloading uses $I_\phi=o_{\rm prev}+K_pe$, with $o_{\rm prev}$ the prior drift request in BOTH or trim in PHASE. Branch changes remap these meanings. During holdover, trim is held and clock integrators stop. Warm recovery uses a five-sample drift check and repeats acquisition if its 0.05 ppb residual-mean criterion is not met. Recovery captures a fresh local phase reference while preserving the delivered trim. Relative-lock indication and permission to steer are separate.

\FloatBarrier
\appendix{Thermal startup, regulation and duty centering}\label{app:thermal}
\subappendix{Initial temperature goal}\label{app:thermal-startup}
Once temperature and OCXO readback are valid, we apply 100\% heater duty. After 60 s, we fit temperature slope over a rolling 30 s window. When the slope stays below approximately $0.4\,^\circ\mathrm{C}\,\mathrm{min}^{-1}$ for 20 continuous seconds, we capture the current raw temperature as the initial goal, clipped to 0--60$^\circ$C. We then request zero heater duty and start the inner PID with zero integral; it builds the required holding duty.

The measured effective heat capacity of the PCB and aluminium block together is $C_{\rm eff}\simeq75$ J K$^{-1}$. At full heater power $P_{\rm on}=1$ W, stored heat increases at $C_{\rm eff}\dot T_{\rm warm}$. For unchanged ambient conditions and self-heating, the remaining power balances net heat loss, so the external power needed to hold the capture temperature is
\begin{equation}
 P_{\rm hold}\simeq P_{\rm on}-C_{\rm eff}\dot T_{\rm warm}
 \simeq 1\,\mathrm W-(75\,\mathrm{J\,K^{-1}})
 \left(\frac{0.4}{60}\,\mathrm{K\,s^{-1}}\right)
 =0.5\,\mathrm W.
 \label{eq:startup-holding-power}
\end{equation}
Thus capture near this warming rate targets approximately 50\% holding duty. Two thermal fits gave corresponding thresholds of 0.423 and 0.374$^\circ$C min$^{-1}$, motivating the rounded 0.4$^\circ$C min$^{-1}$ choice.

\subappendix{Inner PID settings}\label{app:thermal-inner}
We fitted the thermal plant from heater-pulse recordings, assuming 1 W full heater power. The fits gave effective heat capacity 71--80 J K$^{-1}$, thermal conductance $G_{\rm th}=0.148$--0.190 W K$^{-1}$ and delay 6--7 s. For an initial proportional-gain estimate, we used a desired response scale $T_r=20$ s, representative delay $L=6.5$ s and measurement-filter time $T_m=5$ s:
\begin{equation}
 K_{P,T}^{(0)}\approx\frac{C_{\rm eff}}{P_{\rm on}(T_r+L+T_m)}
 =\frac{71\,\mathrm{J\,K^{-1}}}{(1\,\mathrm W)(31.5\,\mathrm s)}
 \simeq2.25\,\mathrm{K^{-1}}.
 \label{eq:thermal-initial-kp}
\end{equation}
Offline model comparisons led to $K_{P,T}=2.6$ K$^{-1}$. We selected $K_{I,T}=0.016$ (K s)$^{-1}$ to balance disturbance rejection and overshoot, giving $T_i=K_{P,T}/K_{I,T}=162.5$ s. Derivative anticipation with $T_d=6.5$ s gives $K_{D,T}=K_{P,T}T_d=16.9$ s K$^{-1}$.

The simulations included the actual fixed-point controller, sensor quantization, delay, saturation and several thermal models. We tuned for faster disturbance recovery; the final values are model-guided engineering choices. The implemented parallel PID uses a separately filtered derivative of the raw temperature measurement \citep[Chap.~10]{AstromMurray2008}; duty is a dimensionless fraction, with 1 denoting full heater power.
\begin{center}
\small
\begin{tabular}{@{}lll@{}}
\toprule
Parameter & Value & Unit \\
\midrule
$K_{P,T}$ & 2.6 & K$^{-1}$ \\
$K_{I,T}$ & 0.016 & (K s)$^{-1}$ \\
$K_{D,T}$ & 16.9 & s K$^{-1}$ \\
Proportional/integral temperature-filter time & 5 & s \\
Derivative-filter time & 3 & s \\
Heater frame duration & 1 & s \\
\bottomrule
\end{tabular}
\end{center}
Thermal integration uses elapsed time between fresh updates, normally about 1 s. The inner reference ramps at approximately 0.005 K/s. A 3 s derivative filter and $\pm0.3$ duty contribution limit bound sensitivity to sensor quantization. Total duty is limited to $[0,1]$, rounded to 0.001 and uses anti-windup. Heater frames are 1 s with 1 ms duty resolution. Freshness watchdogs and a 70$^\circ$C thermal-trip threshold bound heater operation.

\subappendix{Slow goal adaptation}\label{app:thermal-outer}
At 1 W, the fitted conductance implies 5.3--6.8 K of heater-induced temperature adjustment, approximately 6 K. We adapt the temperature goal to ambient changes within this finite heater range, keeping actual duty near 50\% to retain both upward and downward power margin. GNSS disciplining is intended to track the associated slow frequency drift while the inner PID suppresses faster thermal fluctuations.

The outer integrator uses filtered actual duty $a_f$, with a 100 s filter; a separate 30 s window qualifies duty history. For duty error $\epsilon=0.5-a_f$, we use the continuous deadband $\mathcal D(\epsilon)=\operatorname{sgn}(\epsilon)\max(|\epsilon|-0.01,0)$. A higher temperature goal requires greater holding duty, so the negative-feedback update is
\begin{equation}
 \dot T_{\rm goal}=\operatorname{clip}
 \left[k_o\mathcal D(0.5-a_f),-r_T,+r_T\right],
 \label{eq:duty-centering}
\end{equation}
with $k_o=0.01$ K s$^{-1}$ per unit duty and $r_T=0.004$ K s$^{-1}$. Thus low duty raises the goal and high duty lowers it; within 49--51\% duty the goal holds. Goals remain within 0--60$^\circ$C. Goal adaptation starts after boot capture and a fresh duty-history window, independently of clock state and GNSS availability. It continues during clock holdover. It pauses when the magnitude of the thermal tracking error exceeds 0.15 K for 10 s and resumes after the error remains below 0.075 K for 30 s; inner regulation continues subject to its own safety conditions.

\FloatBarrier
\appendix{Measurement model and stability analysis}\label{app:analysis}
We distinguish physical output-phase measurements, receiver-reported bias/drift and the reconstructed residuals defined in Appendix~\ref{sec:noise-separation}.

\subappendix{Estimators and processing choices}
We use overlapping Allan deviation (ADEV) and time deviation (TDEV), with the standard estimators of \citet{Allan1966} and \citet[Secs.~5.2.4--5.2.6]{Riley2008}. Phase samples are in seconds, frequency samples are dimensionless, and averaging time is $\tau=m\tau_0$ for recorded sample interval $\tau_0$ and integer averaging factor $m$. TDEV is related to modified Allan deviation (MDEV), which includes additional phase averaging, by $\sigma_x(\tau)=\tau\,\mathrm{MDEV}(\tau)/\sqrt3$. Here $\sigma_x$ is TDEV and $\sigma_y$ denotes ADEV. Relating $\tau\sigma_y$ to timing error requires a specified noise model.

The integrated drift-residual series $Z_{ij}$ is defined in Appendix~\ref{sec:noise-separation}. Interpreting each $D_{ij,n}$ as an interval-average frequency requires compatible receiver and phase-difference gates; logging cadence and the receiver's effective gate or filtering are separate properties. Its TDEV therefore characterizes the integrated estimator, distinct from physical output-phase TDEV or bias-residual TDEV.

The 1 s alignment, mode selection and segment boundaries are described in Sec.~\ref{sec:threeclock}; the component studies alone apply OCXO-only thermal exclusions (Appendix~\ref{app:components}). Individual-clock estimates require matching windows and filters across all three pair statistics. The available information at long averaging times depends on segment lengths and exclusions as well as elapsed span.

\subappendix{Figure uncertainty conventions}\label{app:plot-uncertainty}
The error bars on the primary measured-output curves in Sec.~\ref{sec:campaign-exports}, and on the primary pair-equivalent component and offline-simulation curves in Appendix~\ref{app:component-curves}, have half-width $E(\tau)=s(\tau)\sqrt{\tau/T}$, where $T$ is the recording span and $s$ the plotted deviation. Scaling a pair curve and its bar by $1/\sqrt2$ preserves relative bar size, without establishing source independence. These are record-length guides with no calibrated coverage probability; gaps and exclusions can reduce information further.

Statistical intervals require noise identification and estimator-specific effective degrees of freedom \citep[Secs.~5.3--5.5]{Riley2008}. Overlapping estimator terms are statistically dependent. Confidence intervals for individual-clock estimates from the three-cornered-hat method must also account for covariance between the three estimated pair variances \citep{Ekstrom2006}. We use the error bars solely as record-length guides; calibrated statistical intervals, common systematic errors and simulation-model uncertainty require separate assessment.

\subappendix{Cross-phase readout and common sampling errors}\label{app:readout}
For complex channel amplitudes $z_k$ estimated over common time blocks, the cross-phase measurement model is
\begin{equation}
 \mathcal C_{kl}=\langle z_kz_l^*\rangle,\qquad
 \widehat x_{kl}=\frac{\operatorname{unwrap}\arg \mathcal C_{kl}}{2\pi f_r},
 \qquad f_r=390\,\mathrm{MHz}.
 \label{eq:crossphase}
\end{equation}
For quasi-constant amplitude and phase within a block, $z_k\propto e^{i\phi_k}$ gives the phase difference. To expose common sampling error, write measured channel phase as
\begin{equation}
 q_k=\phi_{i(k)}-\phi_s+n_k,\qquad
 q_k-q_l=\phi_{i(k)}-\phi_{i(l)}+n_k-n_l,
\end{equation}
where $\phi_s$ is common sampling-time error expressed as phase at the selected tone $f_r$, and $n_k$ contains channel/path error; all phases are in radians. Here $i(1)=i(4)=A$, $i(2)=B$ and $i(3)=C$. Thus $q_1-q_4=n_1-n_4$ assesses differential splitter/cable/readout noise while cancelling A and any path error common to both channels. Frequency mismatch, unequal delays and channel timing can impair common-sampling cancellation.

Shared channels correlate pair estimates and reconstructed residuals. If pairs are formed by subtracting the same channel-phase estimates, pair closure provides an algebraic consistency check. The measured duplicated-input floor is reported in Sec.~\ref{sec:threeclock}; it characterizes differential readout/path noise in that configuration, with sensitivity to the difference between the tested paths.

\subappendix{Covariance and three-cornered-hat interpretation}\label{app:covariance}
Let $V_i$ and $C_{ij}$ be individual-clock variance and clock covariance after the same stability-statistic filtering, using identical windows. If pair readout error is uncorrelated with the clock difference, its variance contribution $N_{ij}$ gives
\begin{equation}
 V_{ij}=V_i+V_j-2C_{ij}+N_{ij}.
 \label{eq:correlated}
\end{equation}
Mixed clock--readout terms must also be retained when that assumption fails. Substitution into the cyclic forms of Eq.~\ref{eq:tch} gives
\begin{align}
 \widehat V_B&=V_B-C_{AB}-C_{BC}+C_{AC}
 +\frac{N_{AB}+N_{BC}-N_{AC}}{2},\nonumber\\
 \widehat V_C&=V_C-C_{AC}-C_{BC}+C_{AB}
 +\frac{N_{AC}+N_{BC}-N_{AB}}{2}.
 \label{eq:tch-bc-covariance}
\end{align}
A supplies a third noisy comparison clock. Unequal A--B and A--C covariances redistribute the B/C estimates in opposite directions. Their sum still equals $V_{BC}$ for common processing, as used in Sec.~\ref{sec:campaign-exports}; B--C covariance and readout error remain. Individual estimates are conditional on those terms, and negative variance estimates give gaps in the deviation curves. Individual-clock decomposition therefore requires a specified covariance model \citep{Premoli1993}.

For either plotted deviation, let $V_{BC}=s_{BC}^2$ and let $\widehat V_B,\widehat V_C$ be the corresponding TCH variance estimates. Adding the cyclic expressions in Eq.~\ref{eq:tch}, with identical windows and filters, gives
\begin{equation}
 s_{\rm eq}^2=\frac{V_{BC}}{2}=\frac{\widehat V_B+\widehat V_C}{2}.
 \label{eq:pair-equivalent}
\end{equation}
Thus the pair-equivalent curve is the root-mean-square of the B/C TCH deviations where both estimates are positive. This identity explains why the average does not depend on A; it does not remove B--C covariance or readout noise.

For used navigation-satellite sets $S_B,S_C$, the overlap fractions are $O_B=|S_B\cap S_C|/|S_B|$ and $O_C=|S_B\cap S_C|/|S_C|$. These quantify the configured satellite overlap; its connection to noise covariance is interpreted through the model above. At fixed individual/readout variances, reducing positive $C_{BC}$ increases pair variance; masks also change count and geometry. Shared satellite/correction, environmental and readout errors can cancel from pairs with external discipline enabled or disabled. The method quantifies differential clock agreement, while common errors and absolute UTC epoch accuracy can be assessed through a separate traceable time comparison.

\FloatBarrier
\appendix{Short-run component studies}\label{app:components}
The initial rooftop recordings, approximately 4.5--7.6 h across the five configurations, hold the oscillator frequency word fixed with disciplining disabled. They separate free-oscillator motion from receiver-estimator noise and identify useful crossover ranges for feedback. The main campaign in Sec.~\ref{sec:campaign-exports} measures the disciplined outputs directly over longer records. The confined-sky site tests are reported in Sec.~\ref{sec:confined-sky}.

\subappendix{Scope of the discipline-off measurements}
Initial OCXO characterization is limited to $\tau\leq1000$ s, with thermal-controller state retained for interpretation. Intervals around heater temperature-goal changes are excluded from the OCXO curves only. Receiver-residual statistics extend further where the records support them. In the short unmasked PPP/HAS record, B intermittently lost PPP/HAS because of a loose antenna cable; that comparison uses A--C/$\sqrt2$ without TCH estimates.

The short records guide crossover selection; the long discipline-on measurements assess sustained delivered-output stability without reconstructing the output from receiver observables.

\subappendix{Separating oscillator and receiver-estimator noise}\label{sec:noise-separation}
For unit $i$, let $x_i$ be oscillator time error (s), $y_i=\dot x_i$ fractional frequency, $b_i$ receiver bias (s) and $d_i$ fractional drift. Define equivalent receiver-estimation errors
\begin{equation}
 g_i=b_i-x_i,\qquad h_i=d_i-y_i.\label{eq:receiver-error}
\end{equation}
These include GNSS and receiver-estimator contributions, without assuming $\dot g_i=h_i$. The receiver uses the measured OCXO as its external reference. With consistent time systems, pair ordering and epoch alignment, the oscillator contribution cancels in
\begin{align}
 G_{ij}&=(b_i-b_j)-x_{ij},\nonumber\\
 D_{ij}&=(d_i-d_j)-y_{ij},\label{eq:gnss-residuals}
\end{align}
where $y_{ij}$ is estimated from $x_{ij}$ on the common grid. The residuals are $g_i-g_j$ (s) and $h_i-h_j$ (dimensionless), including readout error. The drift residual is integrated within each segment as $Z_{ij,n+1}=Z_{ij,n}+\Delta t\,D_{ij,n}$, with $\Delta t=1$ s and an arbitrary initial time offset. The independent RFSoC phase measurement supplies the oscillator contribution needed for this separation.

We compute ADEV and time deviation (TDEV), denoted $\sigma_x(\tau)$, for the measured OCXO difference $x_{ij}$, integrated drift residual $Z_{ij}$ and bias residual $G_{ij}$, together with the offline predicted disciplined output. Integrated-drift TDEV characterizes the reconstructed receiver-estimator residual. Appendix~\ref{app:analysis} describes readout error and covariance in these reconstructed quantities.

\begin{figure}[!htbp]
\centering
\vbox{\offinterlineskip
\hbox{\includegraphics[width=\linewidth,trim=0 254.8bp 0 0,clip]{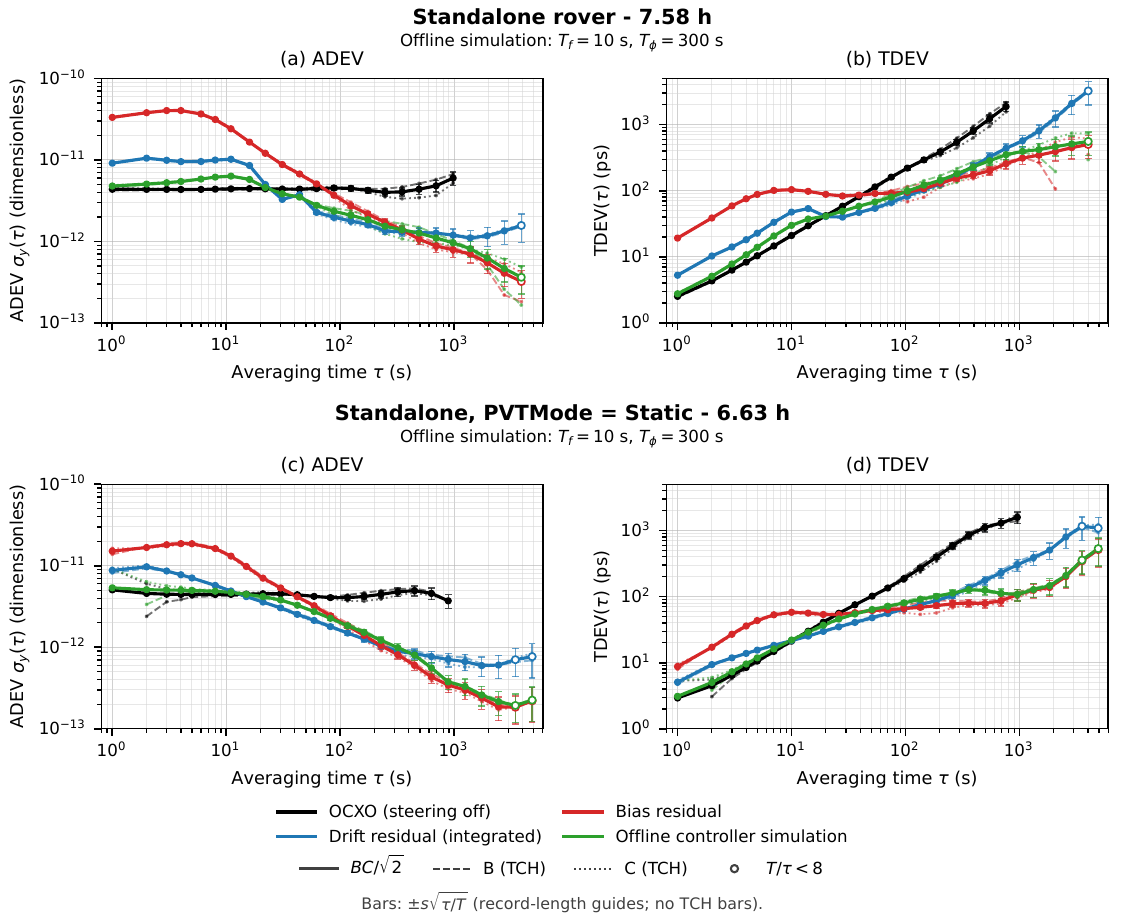}}
\hbox{\includegraphics[width=\linewidth,trim=0 0 0 188bp,clip]{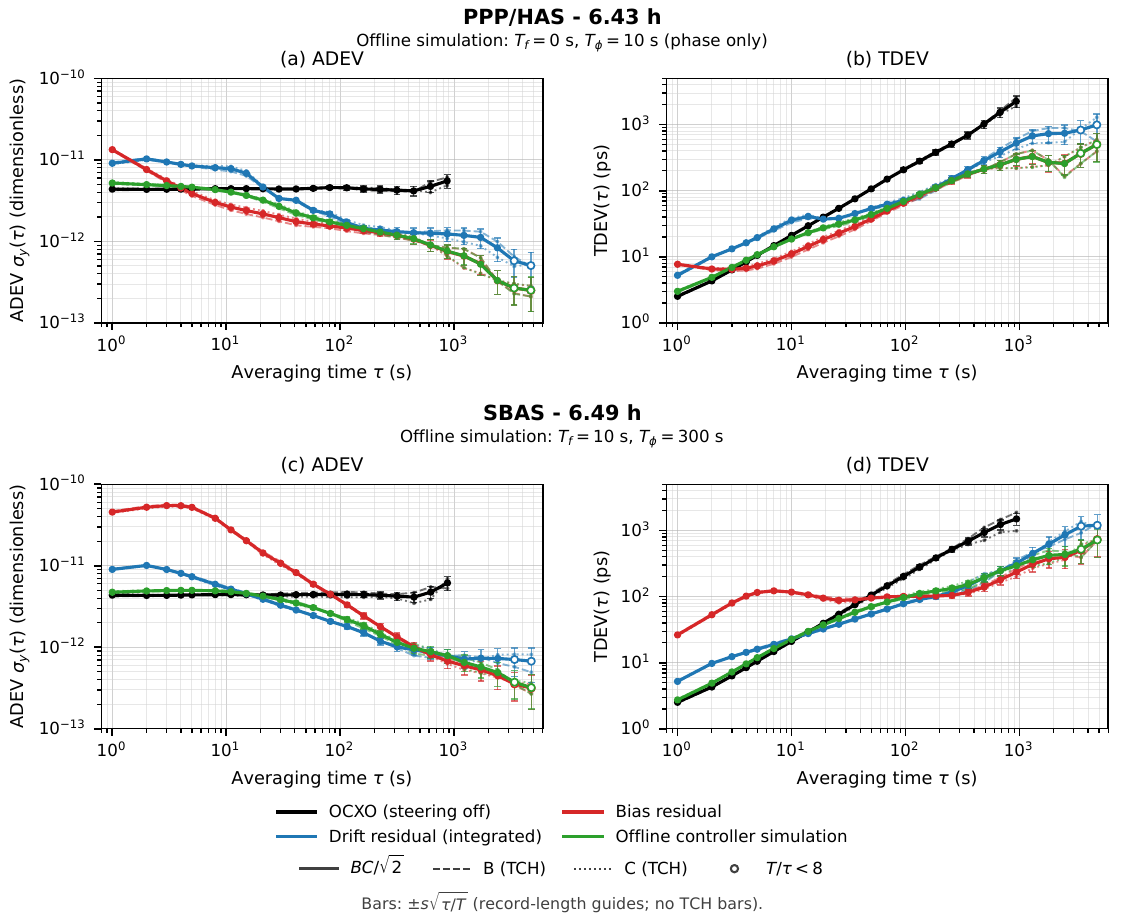}}}
\caption{Rooftop component stability with imposed navigation-satellite exclusions and oscillator steering disabled. Top: standalone rover (7.58 h); bottom: SBAS (6.49 h). Left: ADEV; right: TDEV. Solid curves are B--C/$\sqrt{2}$; dashed/dotted curves are B/C TCH estimates. Black: OCXO differences; blue: integrated drift residuals; red: bias residuals; green: offline controller simulations. Both replays use $T_f=10$ s and $T_\phi=300$ s. OCXO curves stop at 1000 s; other curves extend to 5000 s. Error error bars $\pm s_{\rm eq}\sqrt{\tau/T}$ are uncalibrated record-length guides; hollow markers indicate $T/\tau<8$ (Appendix~\ref{app:plot-uncertainty}).}\label{fig:component-rover-sbas}
\end{figure}

\begin{figure}[!htbp]
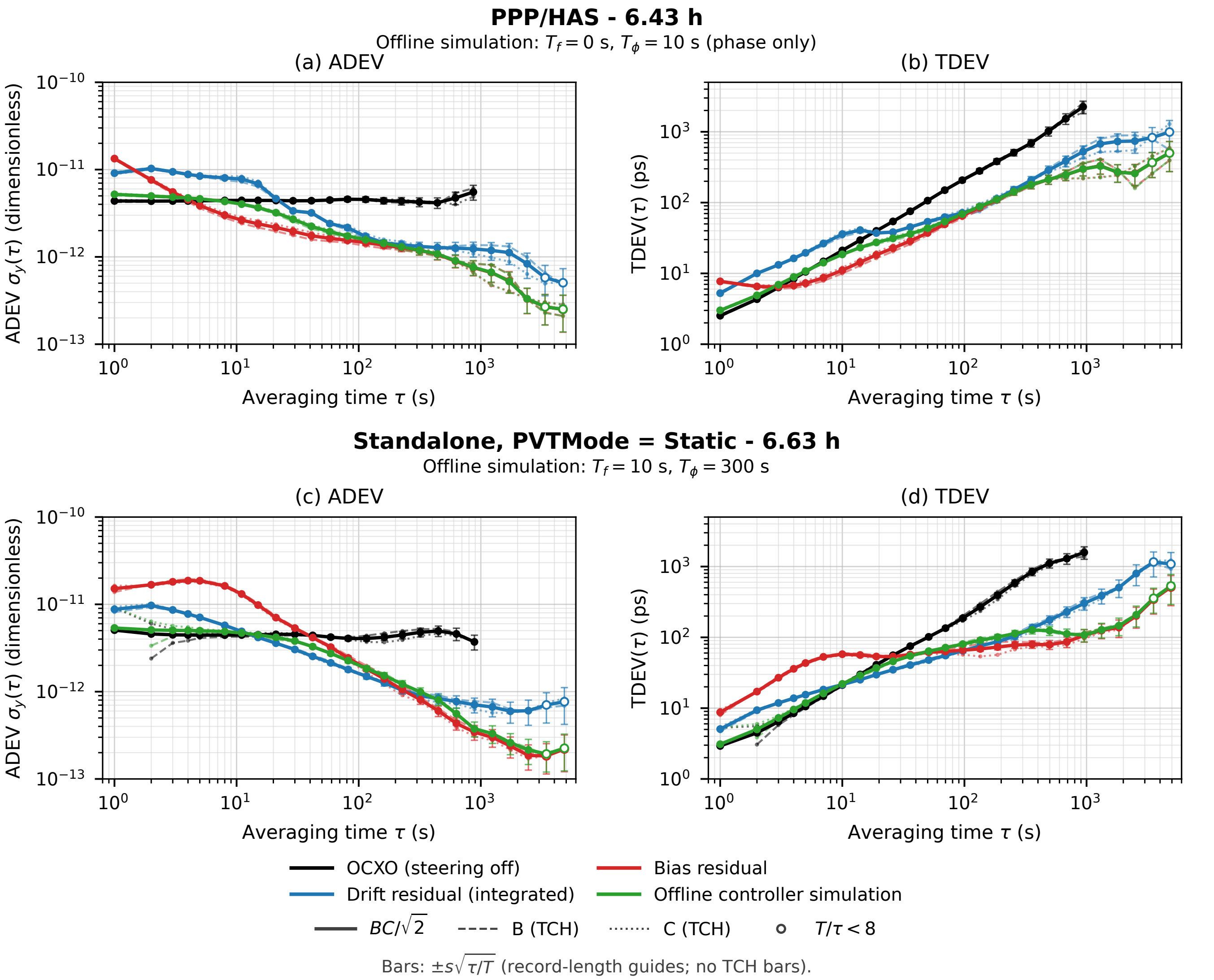

\centering
\vbox{\offinterlineskip
\hbox{\includegraphics[width=\linewidth,trim=0 254.8bp 0 0,clip]{figures/campaign_corrected_modes.pdf}}
\hbox{\includegraphics[width=\linewidth,trim=0 0 0 188bp,clip]{figures/campaign_standalone_modes.pdf}}}
\caption{Rooftop component stability with imposed navigation-satellite exclusions and oscillator steering disabled. Top: PPP/HAS (6.43 h); bottom: standalone \texttt{PVTMode = Static} (6.63 h). Left: ADEV; right: TDEV. Solid curves are B--C/$\sqrt{2}$; dashed/dotted curves are B/C TCH estimates. Black: OCXO differences; blue: integrated drift residuals; red: bias residuals; green: offline controller simulations. PPP/HAS uses phase-only replay with $T_\phi=10$ s; standalone \texttt{PVTMode = Static} uses $T_f=10$ s and $T_\phi=300$ s. OCXO curves stop at 1000 s; other curves extend to 5000 s. Error error bars $\pm s_{\rm eq}\sqrt{\tau/T}$ are uncalibrated record-length guides; hollow markers indicate $T/\tau<8$ (Appendix~\ref{app:plot-uncertainty}).}\label{fig:component-ppp-static}
\end{figure}

\begin{figure}[!htbp]
\centering
\includegraphics[width=\linewidth]{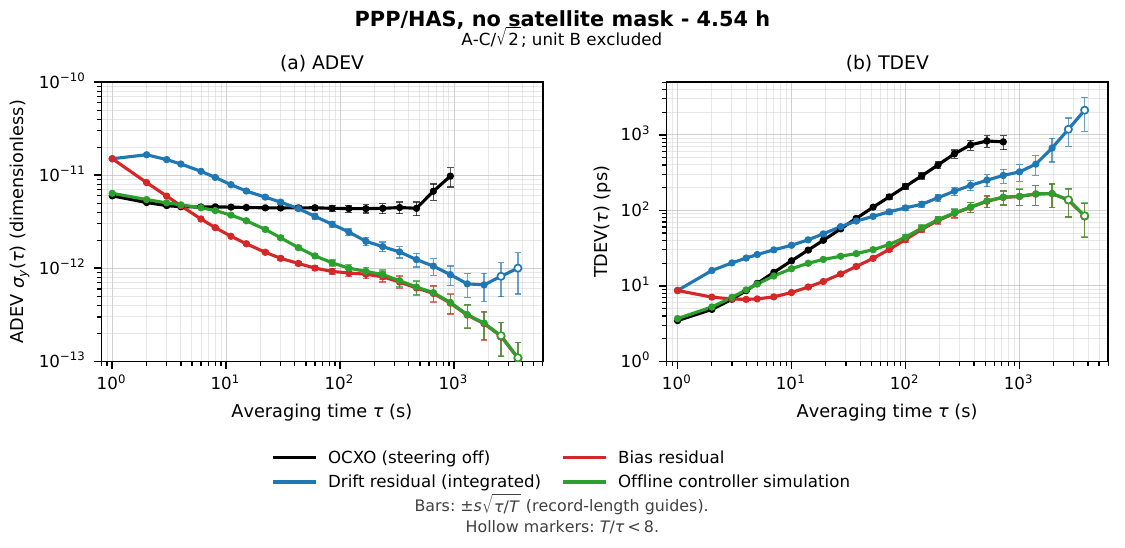}
\caption{Unmasked PPP/HAS stability using A--C/$\sqrt{2}$ with oscillator steering disabled. Left: ADEV; right: TDEV. Colours and axes follow Fig.~\ref{fig:component-rover-sbas}. B is excluded because a loose antenna cable intermittently interrupted its PPP/HAS solution; no TCH estimates are shown. The record spans 4.54 h; error bars are record-length guides and hollow markers indicate $T/\tau<8$.}\label{fig:campaign-mask}
\end{figure}

\subappendix{Component stability and exploratory simulations}\label{app:component-curves}
Figures~\ref{fig:component-rover-sbas}--\ref{fig:campaign-mask} report the component comparisons in the same mode order as Sec.~\ref{sec:campaign-exports}: standalone rover, SBAS, masked PPP/HAS, standalone \texttt{PVTMode = Static}, and unmasked PPP/HAS. The masked standalone-rover, SBAS and PPP/HAS spans are 7.58, 6.49 and 6.43 h, respectively. The standalone \texttt{PVTMode = Static} and unmasked PPP/HAS records span 6.63 and 4.54 h, respectively. Black, blue and red denote measured OCXO differences, integrated drift residuals and bias residuals; green denotes offline controller simulations. Solid curves are the primary pair divided by $\sqrt2$, and broken curves are B/C TCH estimates where available.

The archived PPP/HAS simulation uses phase-only feedback with $T_\phi=10$ s. The archived standalone-rover, SBAS and \texttt{PVTMode = Static} simulations use $T_f=10$ s and $T_\phi=300$ s. These exploratory simulations illustrate the component trade-off. The campaign settings are empirical engineering choices guided by these component studies and separate short discipline-on tuning runs (Table~\ref{tab:campaign-runs}).

At $\tau=10$ s, B--C bias-residual equivalent TDEV is 103.8, 116.8, 11.1 and 57.8 ps for standalone rover, SBAS, masked PPP/HAS and standalone with \texttt{PVTMode = Static}; integrated-drift values are 47.5, 22.9, 35.8 and 21.5 ps, respectively. PPP/HAS thus favours bias feedback at short times, whereas drift supplies the quieter short-to-intermediate-timescale input in the other modes. These diagnostic values characterize reconstructed receiver errors rather than measured disciplined outputs.

Error error bars $\pm s_{\rm eq}\sqrt{\tau/T}$ on the primary pair-equivalent curves, including offline controller simulations, are uncalibrated record-length guides, with $T$ set to each recording span. Hollow markers flag $T/\tau<8$. These guides describe finite record length; they do not quantify simulation-model uncertainty. TCH curves have no error bars. Appendix~\ref{app:plot-uncertainty} explains this convention.

Figure~\ref{fig:campaign-mask} shows PPP/HAS without imposed satellite exclusions, using $s_{AC}/\sqrt{2}$ because of B's intermittent antenna connection. At 10 s, the bias-residual equivalent TDEV is 8.15 ps and the integrated-drift value is 34.5 ps. The masked PPP/HAS result in Fig.~\ref{fig:component-ppp-static} uses B--C; differences between these records therefore include the selected hardware pair and recording conditions as well as satellite count, geometry and covariance.

\FloatBarrier

\FloatBarrier

\bibliographystyle{ws-jai}
\bibliography{references}
\end{document}